\documentclass[useAMS,usenatbib]{mn2e}
\usepackage{epsfig,graphics}

\def\Re{\mbox{$R_{\rm eff}$}}

\def\Msun{\mbox{$M_\odot$}}

\def\mst{\mbox{$M_{\star}$}}

\def\lsim{\mathrel{\rlap{\lower3.5pt\hbox{\hskip0.5pt$\sim$}}
    \raise0.5pt\hbox{$<$}}}                % less than or approx. symbol
\def\gsim{~\rlap{$>$}{\lower 1.0ex\hbox{$\sim$}}}

\def\Msun{\mbox{$M_\odot$}}

\def\lsim{\mathrel{\rlap{\lower3.5pt\hbox{\hskip0.5pt$\sim$}}
    \raise0.5pt\hbox{$<$}}}
\def\gsim{~\rlap{$>$}{\lower 1.0ex\hbox{$\sim$}}}

\def\Re{\mbox{$R_{\rm eff}$}}

\def\mst{\mbox{$M_{*}$}}

\def\gage{\mbox{$\nabla_{\rm age}$}}

\title[Stellar population gradients]{Stellar population gradients from cosmological simulations: dependence on mass and environment in local galaxies}

\author[]{\noindent
C.~Tortora$^{1}$\thanks{E-mail: ctortora@physik.uzh.ch}, A.D. Romeo$^{2}$,
N.R. Napolitano$^{3}$, V. Antonuccio-Delogu$^{4,5}$, A. Meza$^{2}$
\and J. Sommer-Larsen$^{6,7}$, M. Capaccioli $^{8,9}$
\\
~\\
$^1$ Universit$\ddot{a}$t Z$\ddot{u}$rich, Institut f$\ddot{u}$r
Theoretische Physik, Winterthurerstrasse 190, CH-8057,
Z$\ddot{u}$rich, Switzerland\\
$^2$ Universidad Andres Bello,
Departamento de Ciencias
Fisicas, Av. Republica 220, Santiago, Chile\\
$^3$ INAF -- Osservatorio Astronomico di Capodimonte, Salita Moiariello 16, I-80131 - Napoli, Italy\\
$^4$ INAF -- Osservatorio Astrofisico di Catania, Via S. Sofia 78, I-95123 - Catania, Italy \\
$^5$ Scuola Superiore di Catania, Via San Nullo, 5/i, 95123 Catania, Italy\\
$^6$ Excellence Cluster Universe, Technische Universit$\ddot{a}$t M$\ddot{u}$nchen, Boltzmannstr. 2, D-85748 Garching bei M$\ddot{u}$nchen, Germany \\
$^7$ Dark Cosmology Centre, Niels Bohr Institute, University of Copenhagen, Juliane Maries Vej 30, DK-2100 Copenhagen, Denmark\\
$^8$ Dipartimento di Scienze Fisiche, Universit\`{a} di Napoli
Federico II, Compl. Univ. Monte S. Angelo, 80126 - Napoli, Italy\\
$^9$ MECENAS, Universit\`{a} di Napoli Federico II and
Universit\`{a} di Bari, Italy}

\begin{document}
\date{Accepted  Received }
\pagerange{\pageref{firstpage}--\pageref{lastpage}} \pubyear{xxxx}
\maketitle

\label{firstpage}
\begin{abstract}
The age and metallicity gradients for a sample of group and
cluster galaxies from N-body+hydrodynamical simulation are
analyzed in terms of galaxy stellar mass. Dwarf galaxies show null
age gradient with a tail of high and positive values for systems
in groups and cluster outskirts. Massive systems have generally
zero age gradients which turn to positive for the most massive
ones. Metallicity gradients are distributed around zero in dwarf
galaxies and become more negative with mass; massive galaxies have
steeper negative metallicity gradients, but the trend flatten with
mass. In particular, fossil groups are characterized by a tighter
distribution of both age and metallicity gradients. We find a good
agreement with both local observations and independent
simulations. The results are also discussed in terms of the
central age and metallicity, as well as the total colour, specific
star formation and velocity dispersion.
\end{abstract}


\section{Introduction}\label{sec:intro}

\begin{figure*}
\psfig{file=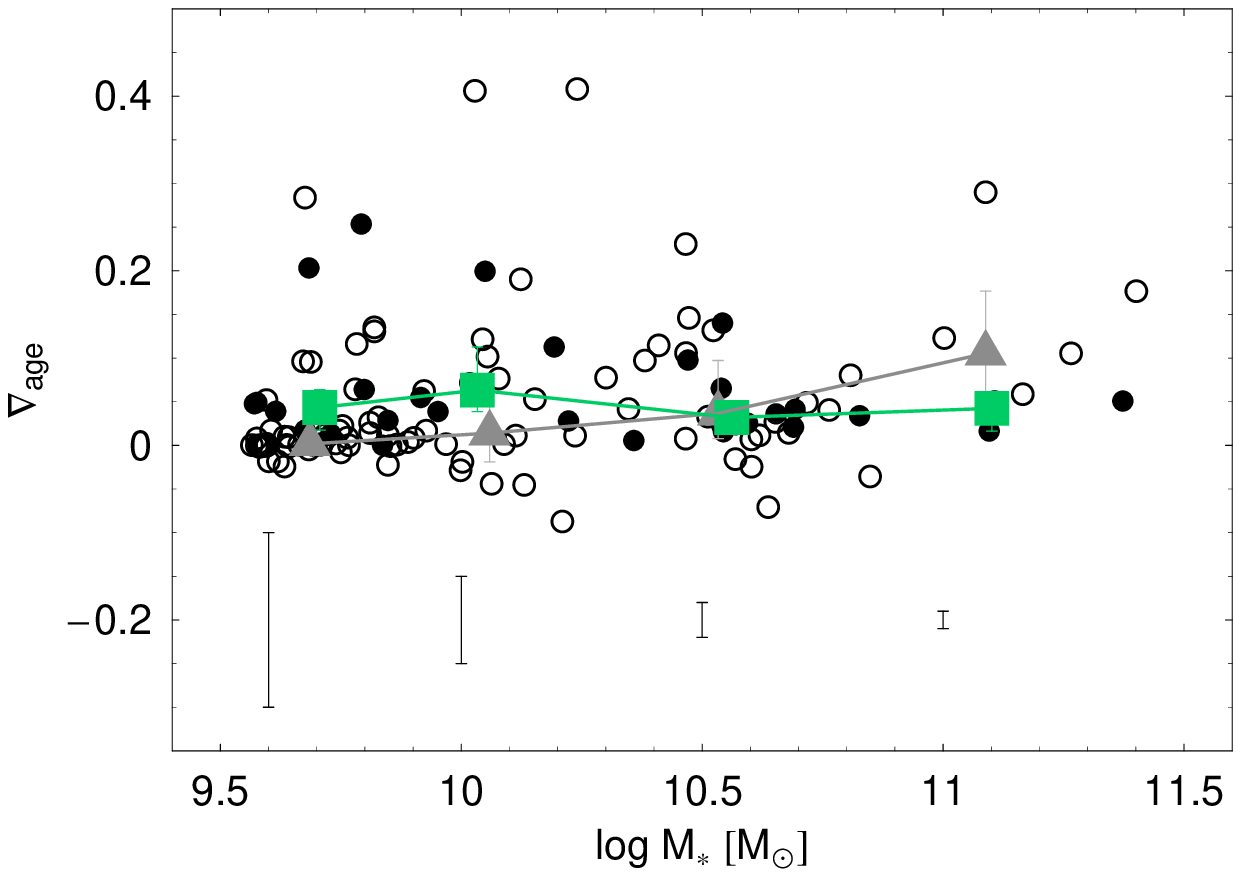, width=0.48\textwidth}
\psfig{file=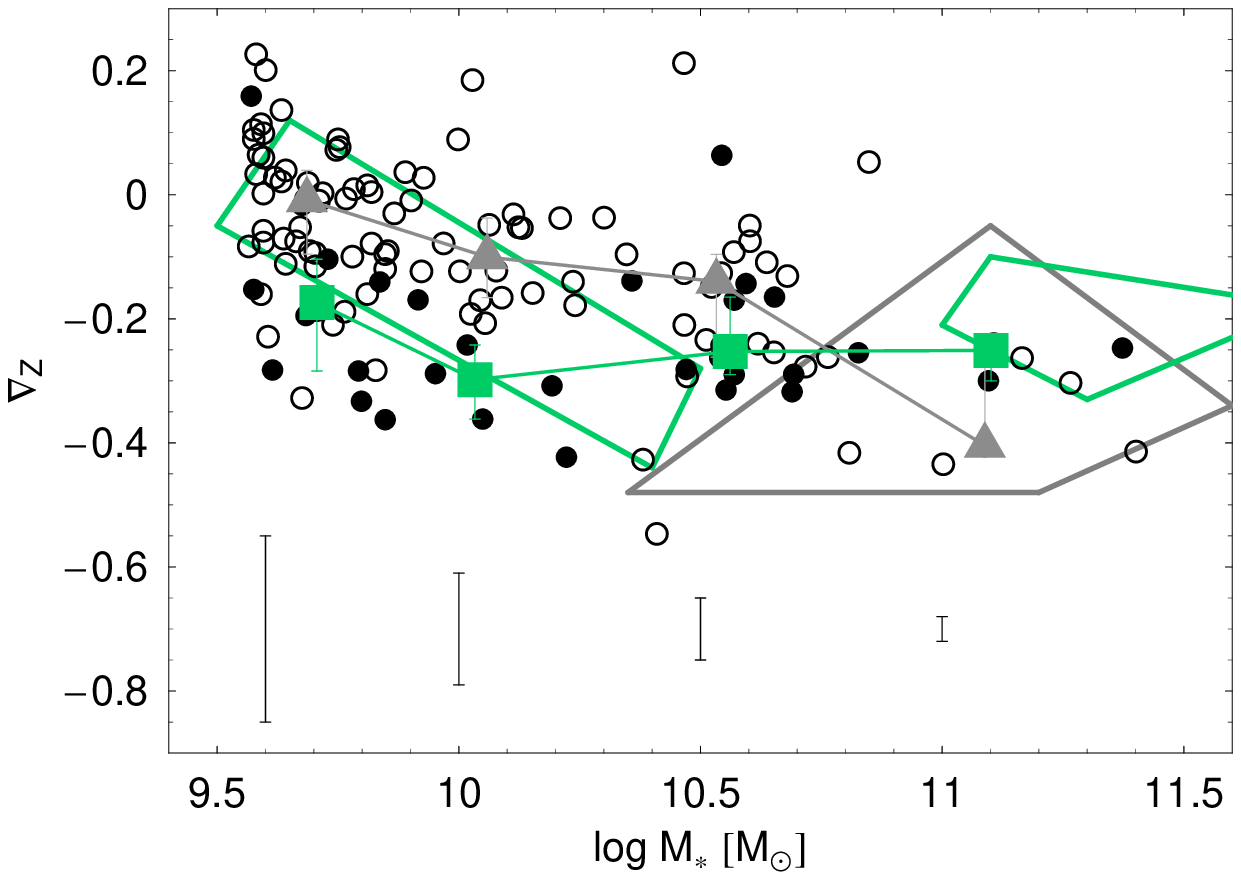, width=0.48\textwidth}\\
\psfig{file=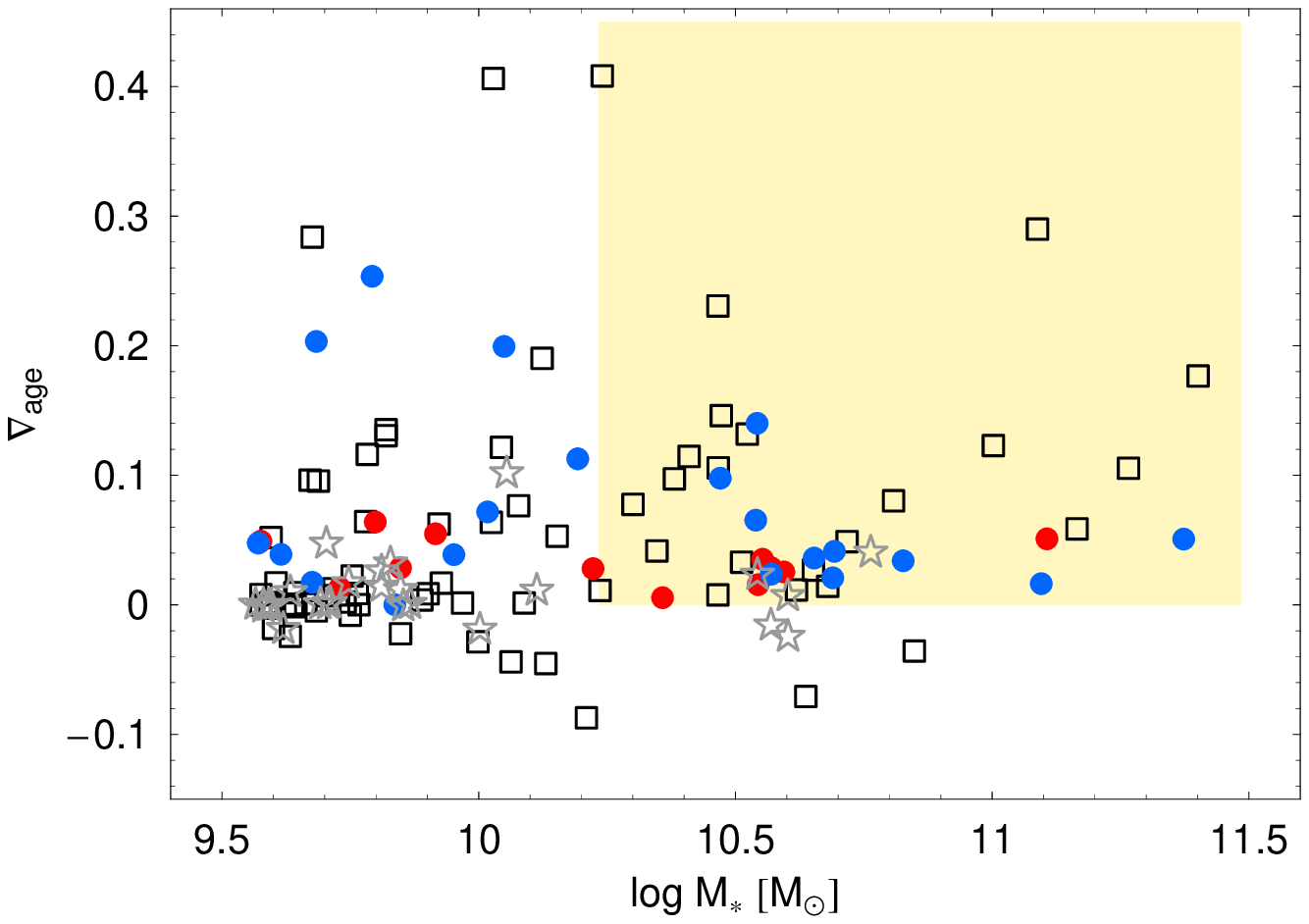, width=0.48\textwidth}
\psfig{file=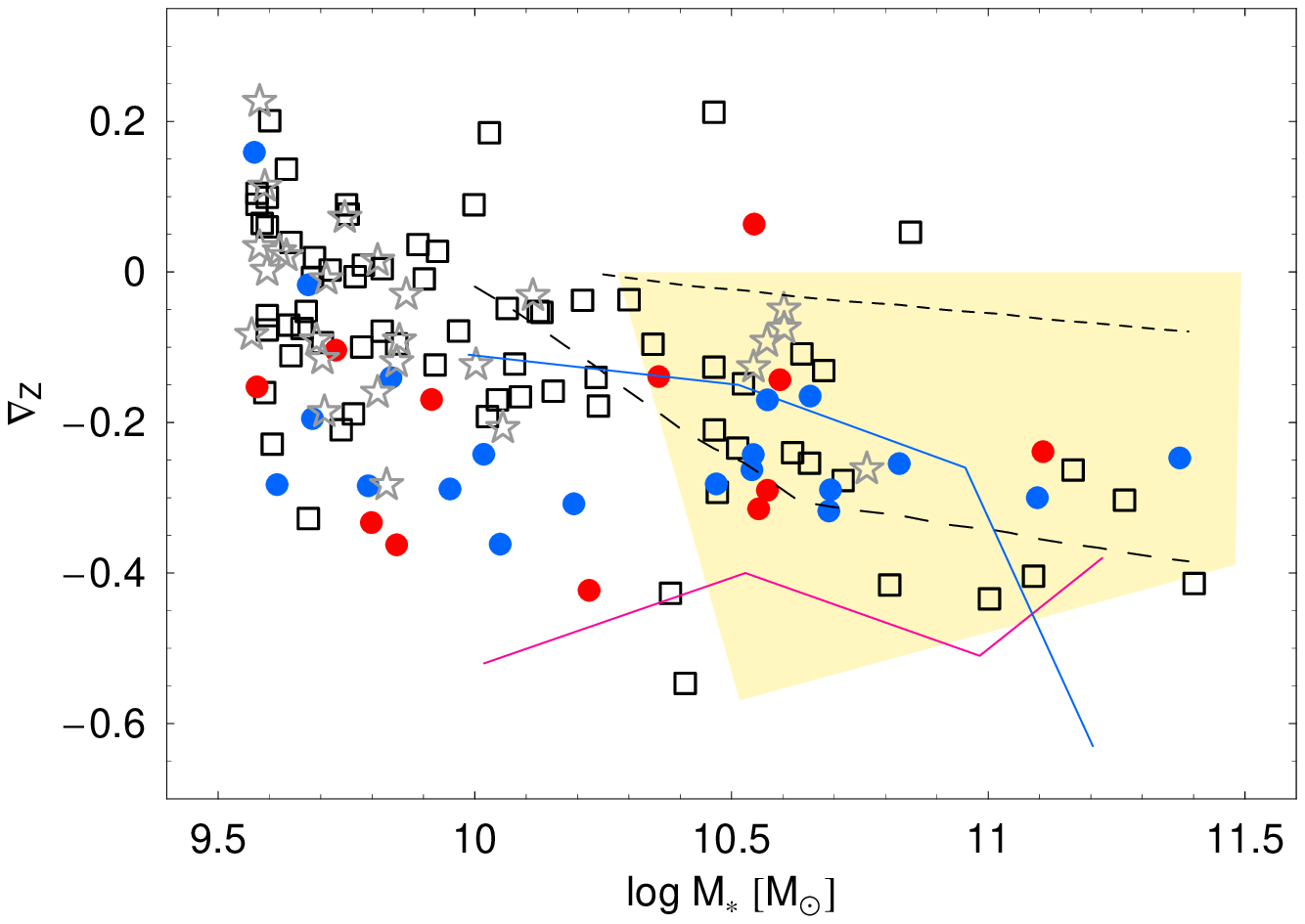, width=0.48\textwidth} \caption{ Age (left
panels) and metallicity (right panels) gradients as a function of
stellar mass (assuming an AY IMF). {\it Top panels.} Black filled
and black open circles are for cluster and group members. The
lines connected by gray triangles and green boxes are the medians
for groups and clusters, respectively. The error bars are the
quartiles of sample distribution in each mass bin. At the bottom
of the panels we also show the uncertainties on galaxies at
different stellar masses, shown as error bars and determined as
described in the text. On the right panel we also show the results
from local cluster (green thick lines) and groups (gray thick
lines) galaxies in \citet{Spolaor09}, the polygonals bound the
regions where these data are distributed. {\it Bottom panels.}
Black open boxes, gray open stars, red and blue points are for
galaxies in normal and fossil groups, inner and external regions
of clusters, respectively. Our results are compared with findings
from other simulations. The predicted metallicity gradients from
dissipative collapse model in \citet{KG03} and those from mergings
in \citet{BS99} are shown as long- and short- dashed lines. The
shaded region is for remnants of major-mergings between gas-rich
disk galaxies in \citet{Hopkins+09a}. Blue and pink lines are the
results from the chemo-dynamical model in \citet{Kawata01},
respectively for strong and weak SN feedback. }
\label{fig:grad_vs_mass_z0}
\end{figure*}

Radial profiles of colours, ages and metallicities of stellar
populations have been shown to be efficient tools to discriminate
among different galaxy formation scenarios across a wide range of
masses and environments. Earlier observational works were unable
to assess the dependence of gradients on mass, mainly because of
the limited samples studied (\citealt{Peletier+90a},
\citealt{Davies+93}, \citealt{KoAr99}, \citealt{TO2003},
\citealt{LaBarbera2005}). Only recently, trends of gradients with
mass have been more firmly assessed (e.g. \citealt{Forbes+05}),
pointing towards different physical mechanisms forging the
behaviour of more and less massive galaxies (\citealt{Spolaor09},
\citealt{Rawle+10}, \citealt{Tortora+10CG}). However, the
dependence of the stellar population gradients on environment is
still controversial (\citealt{T+00}; \citealt{TO2000, TO2003}),
although it might have a non marginal role (e.g.
\citealt{LaBarbera2005}).

Simulations of galaxy formation  are the ideal tool to interpret
observations in terms of the underlying physical mechanisms
driving the evolutive processes of galaxies in various
environments and a large mass range. For instance, in
\cite{Tortora+10CG} (T+10 hereafter) we have compared the stellar
population (age and metallicity) gradients of a wide sample of
local galaxies extracted from the Sloan Digital Sky Survey, with a
number of literature simulations (of both monolithically
collapsing systems and merger remnants). We have confirmed the
metallicity as the main factor in shaping the colour gradient as a
function of mass. However, age gradients are still important,
particularly for younger galaxies. We have shown that less massive
systems, which are unlikely to get assembled by merging, have
experienced a simple monolithic collapse, resulting into a
negative and steep metallicity gradient (e.g. \citealt{Larson74,
Larson75}, \citealt{Carlberg84}) which decreases with mass
(\citealt{KG03}, \citealt{Pipino+10}). Supernovae (SN) feedback is
strong in this mass regime, reaching its maximum efficiency in the
lowest mass systems and this contributes to produce such negative
gradients (\citealt{Pipino08}). On the higher mass side instead,
the trend with mass is inverted, demanding the contribution of
strong joint effect of mergings and AGN feedback (\citealt{BS99},
\citealt{Ko04}, \citealt{Hopkins+09a}, \citealt{Sijacki+07}). In
spite of these results, recently \cite{Pipino+10} have reproduced
the flattening of metallicity gradients in massive systems within
a monolithic scenario of galaxy formation, pointing to a
fundamental role of star formation efficiency in shaping the trend
with mass.

In the present paper we will use the {\tt TreeSPH}
N-body+hydrodynamical simulations described in \cite{Romeo+06} and
\cite{Romeo+08} (hereafter R+06 and R+08) to derive the profiles
of both age and metallicity inside the galaxies and discuss them
as a function of stellar mass and environment. This smoothed
particle hydrodynamics (SPH) simulation has some advantages when
compared with either semi-analytical models (SAM) or pure
hydrodynamical simulations of individual objects. In general SAMs
make use of large scale N-body simulations and rely upon different
sets of assumptions and approximations (e.g. dark matter haloes
are spherically symmetric and infalling gas is shock-heated to the
virial temperature of the halo) that depend on many interconnected
and variable parameters (such as star formation and feedback
efficiency or IMF). In this scheme, the evolution of galaxies is
adjusted on top of the pure collisionless DM haloes, by means of
merger trees and models of synthetic colours. On the other hand
SPH simulations contain fewer simplifying assumptions (as, for
example, no restrictions on halo geometry), but have to restrict
themselves on smaller scales (namely, galactic) in order to
maintain a high resolution, resulting in a more limited dynamical
range. However, in terms of gas cooling both approaches are
confirmed to give similar results (\citealt{Benson+01,Helly+03}).
Our approach roughly lies in between, since the formation of
galaxies is followed {\it ab initio} within a large cosmological
volume, thus accounting for their interaction with a large scale
environment. At the same time, it allows for describing in a
self-consistent way the mutual cycle between inter-galactic
medium, star formation and stellar feedback at the cluster scale,
by means of the baryonic physics implemented in the hydrodynamical
code, in particular the chemical enrichment of the gas surrounding
galaxies (for details, see R+06). This is though attained at
expenses of the final stellar resolution, that cannot reach the
level of either pure N-body simulations used in SAMs, nor of the
single-object hydrodynamical simulations at smaller scale. In this
paper then we will test our SPH simulations against both
observations and models, with the caveat of latter's different
resolution: e.g. those of gas-rich mergers between disks by
\cite{BS99} and \cite{Hopkins+09a}, or the cosmological
simulations including a chemo-dynamical model by \cite{KG03}.

In \S\ref{sec:sim_setup} we will present the simulation setup and
the fitting procedure adopted to recover the gradients;
systematics in the fitting procedure and simulation resolution are
analyzed in \S\ref{app:syst}. In \S\ref{sec:results} and
\ref{sec:discussions} the trends with mass and environment are
analyzed and we give an interpretation of the physical processes,
while the conclusions are drawn in \S\ref{sec:conclusions}.

\section{Simulation setup}\label{sec:sim_setup}

We have extracted simulated galaxies from the SPH simulations in
R+06 of two clusters of temperatures $T \sim 3$ (C1) and $6 \, \rm
keV$ (C2) and 12 groups ($T \sim 1.5 \, \rm keV$), four of which
are fossil. They were drawn and resimulated from a DM-only
cosmological simulation run with the code FLY (\citealt{AD+03}),
for a standard flat $\Lambda$CDM cosmological model ($h = 0.7$,
$\Omega_{m} = 0.3$, $\sigma_{8} = 0.9$) with $150 h^{-1} \, \rm
Mpc$ boxlength. When resimulating with the hydrocode, baryonic
particles were 'added' to the original DM ones, which were split
according to a chosen baryon fraction $f_{b}= 0.12$.

Galaxies are composed by $N_{\rm par}$ star particles bound to the
DM halo; each star particle represents a Single Stellar Population
(SSP) of total stellar mass corresponding to the stellar mass
resolution ($M_{\star, SSP}$) of the simulation. This is
$M_{\star, SSP} = 3.1 \times 10^{7} h^{-1}$ for groups and C1,
while $M_{\star, SSP} = 25 \times 10^{7} h^{-1}$ for C2. The total
luminosity and mass for each galaxy is defined as the sum of the
luminosities and masses of their $N_{par}$ star particles. The
individual stellar masses are distributed according to an
\citet[hereafter AY]{AY87} IMF; each of these SSPs is
characterized by its age and metallicity (Z), from which
luminosities are computed by mass-weighted integration of the
Padova isochrones. The ``standard'' super-wind model for SN is
adopted, i.e. a prescription for SN feedback in which $70\%$ of
the energy feedback from SN type II goes into driving galactic
super-winds (see R+06 for more details). Although AGN feedback is
not taken into account in this simulation, R+08 has shown that the
colour properties of galaxies are fairly good reproduced; however,
AGNs would play a major role only in high mass systems.

Because of the galactic winds expelling baryons out into the IGM,
lower mass galaxies have a higher fraction of DM over stars (and
gas), which also results in a low absolute number of star
particles. Throughout this paper we will only deal with the
stellar component, willing to remain as much as consistent with
the approach followed in the observations. For this reason it is
important to determine a lower limit to the number of star
particles, in order to avoid systems whose stellar component is
not sufficiently resolved.
%
%We have applied a completeness limit in stellar mass of $\log \mst
%> 9.5\, \Msun$, corresponding to $N_{par} > 71$ for the cluster C1
%and the groups, while is $N_{par} > 9$ for the lower resolution
%C2, however for C2 an extra cut at $N_{par} > 15$ has been
%applied. We are left with a sample including 130 galaxies from the
%two clusters and 102 from the 12 groups, at $z$=0. However,
%After these further cut, while the group population is almost
%unchanged, since now includes 97 galaxies, the cluster galaxies
%are enormously reduced to 32 systems.
%
We have originally applied a completeness limit in stellar mass of
$\log \mst > 9.5\, \Msun$, but to be more conservative we have
only retained those systems having $N_{par} > 80$. We are left
with a sample including 32 galaxies from the two clusters and 97
from the 12 groups, at $z$=0. The systematics in the galaxies with
a lower number of particles will be discussed in \S\
\ref{app:app1b}, where we also show how the results could change
when systems with $N_{par}\leq 80$ were included.

Brightest central galaxies (BCG) are excluded from this analysis,
since their diffuse envelope mixing with the intra-cluster light
makes difficult to fit a reliable luminosity profile on the basis
of the galaxy-bound star particles only.

For a given stellar population parameter X ($= age, \, Z$), we
will assume the radial variation as $\log X (R)= a_{X} + b_{X}
\log R$, and define the X gradient as the angular coefficient
$b_{X} = \nabla_{X} = \frac{\delta (\log X)}{\delta \log R}$. For
convenience, we adopt as central quantity the value of $\log X$ at
$R=1 \, \rm kpc$, i.e. the intercept $a_X$. By definition, if the
gradient is positive, i.e. $\nabla_{X}>0$, then X is lower in the
central regions, while X decreases as a function of R when the
gradient is negative, i.e. $\nabla_{X}<0$.

The centre of the galaxy is defined as the median of the positions
of all the star particles. The fit is made after having discarded
those particles with age and metallicities $>3 \sigma$ away from
the median (see \S\ref{app:app1a} for further details). To take
into account disturbed galaxy shapes, we performed the log--log
fit on three projected orthogonal planes and took the median of
the parameters to obtain the final gradient values.

\begin{figure*}
\psfig{file=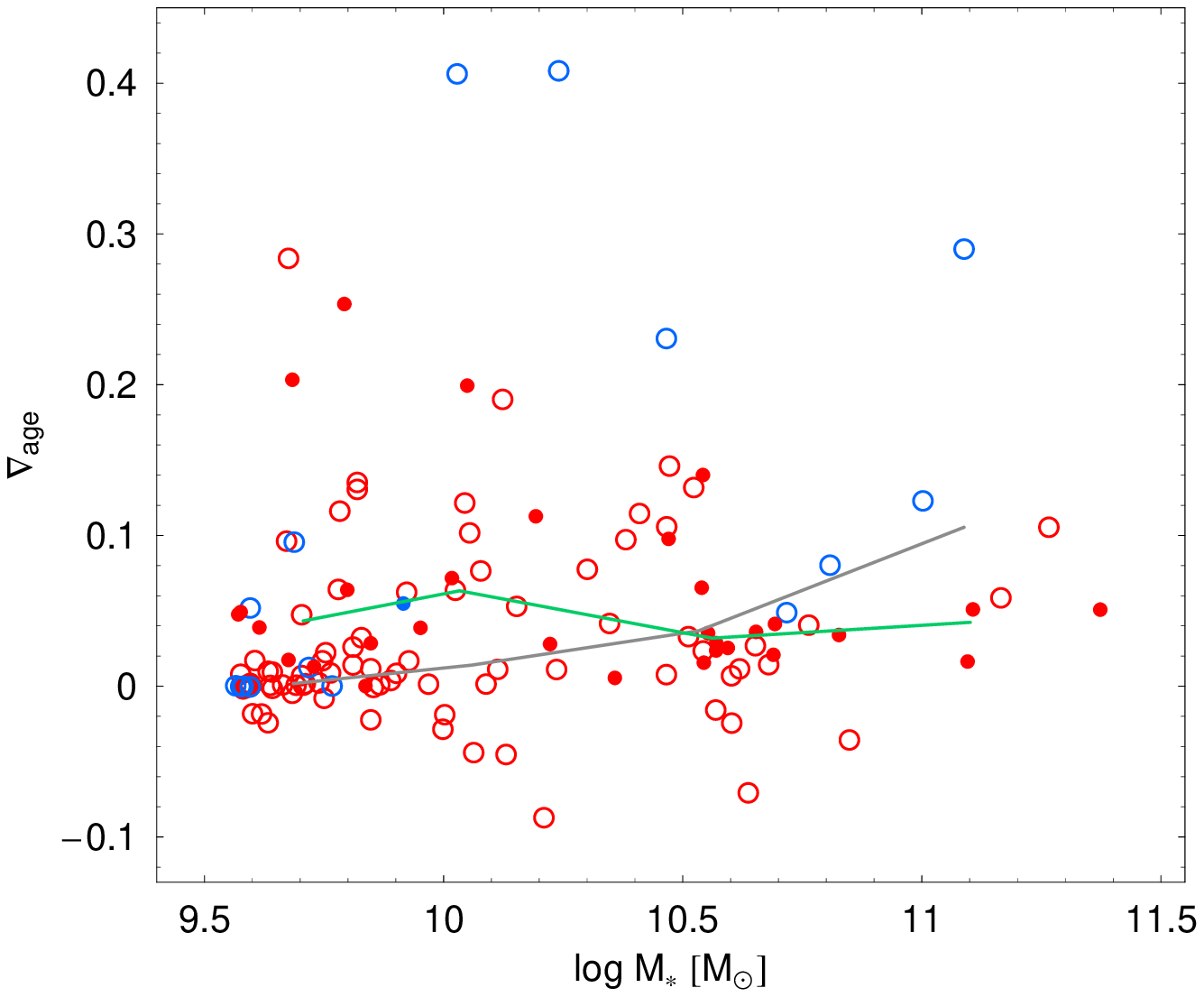, width=0.32\textwidth} \hspace{0.15cm}
\psfig{file=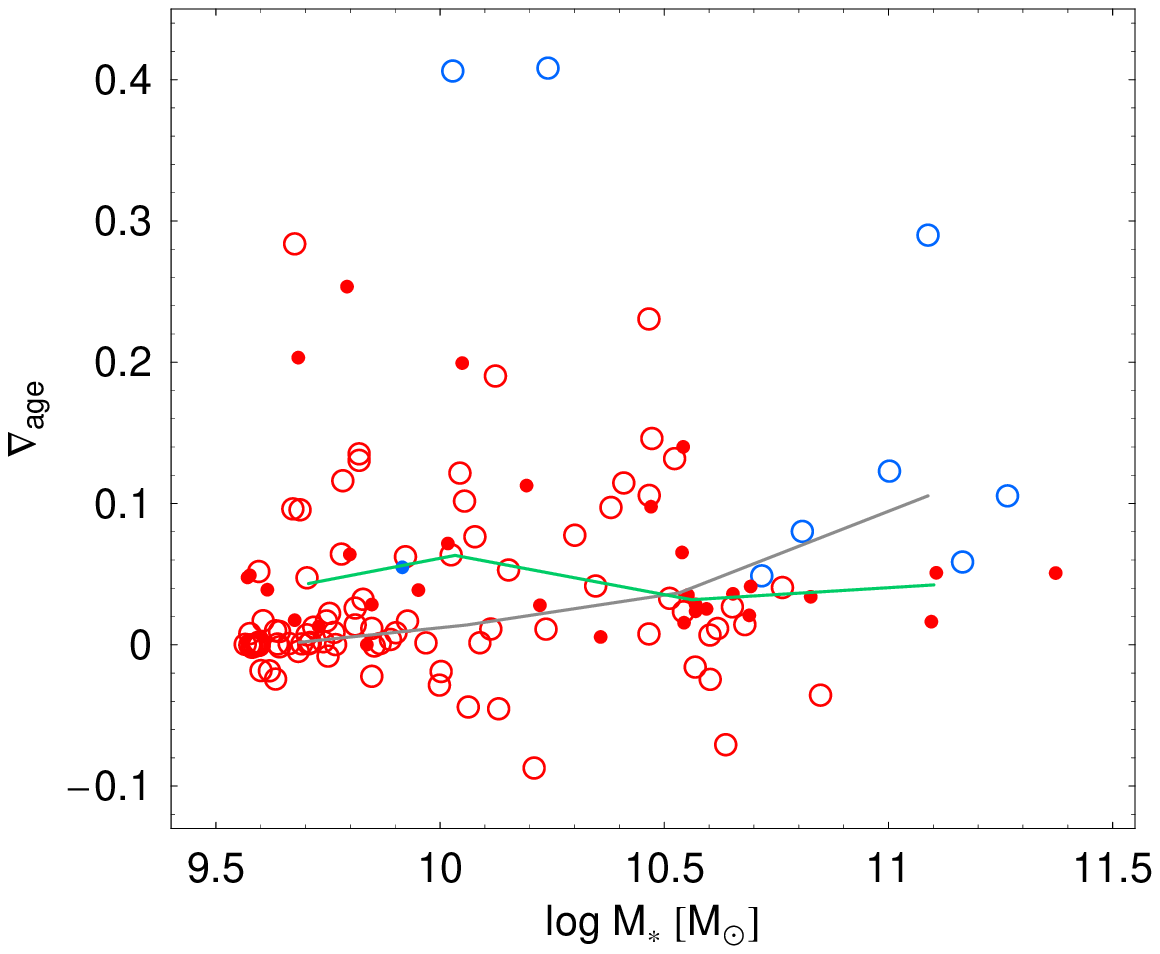, width=0.32\textwidth}\hspace{0.15cm}
\psfig{file=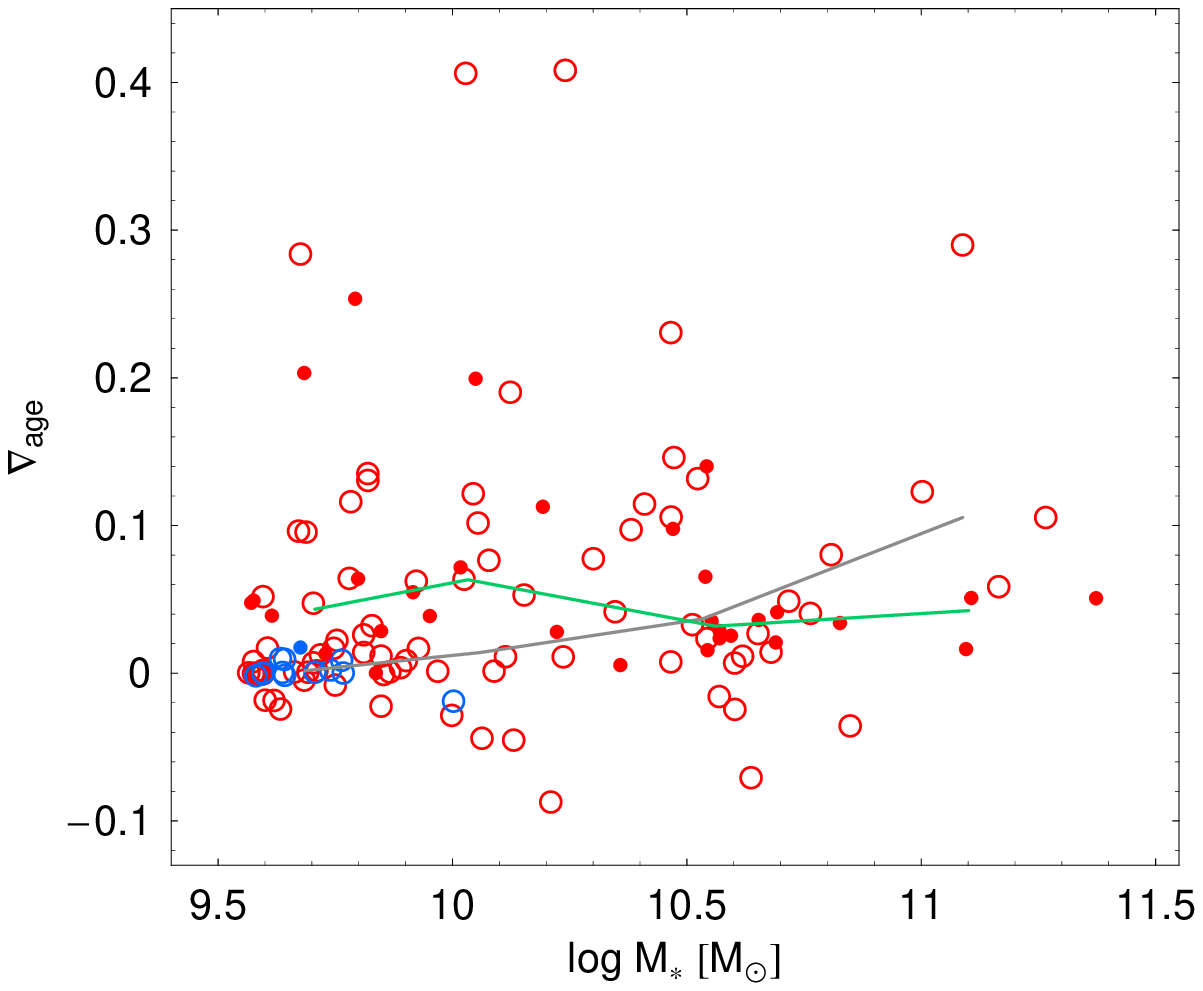, width=0.32\textwidth}\\
\psfig{file=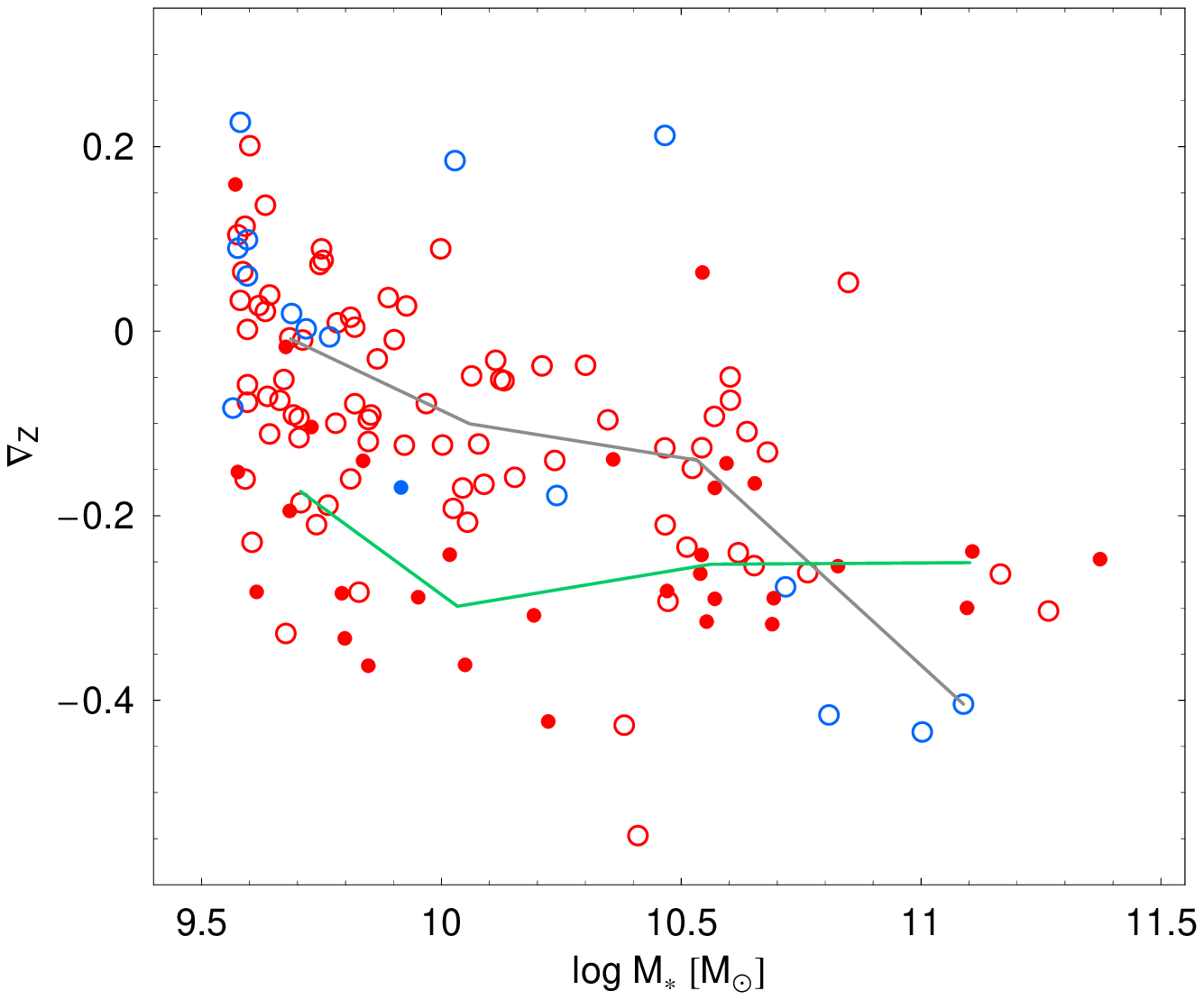, width=0.32\textwidth}\hspace{0.15cm}
\psfig{file=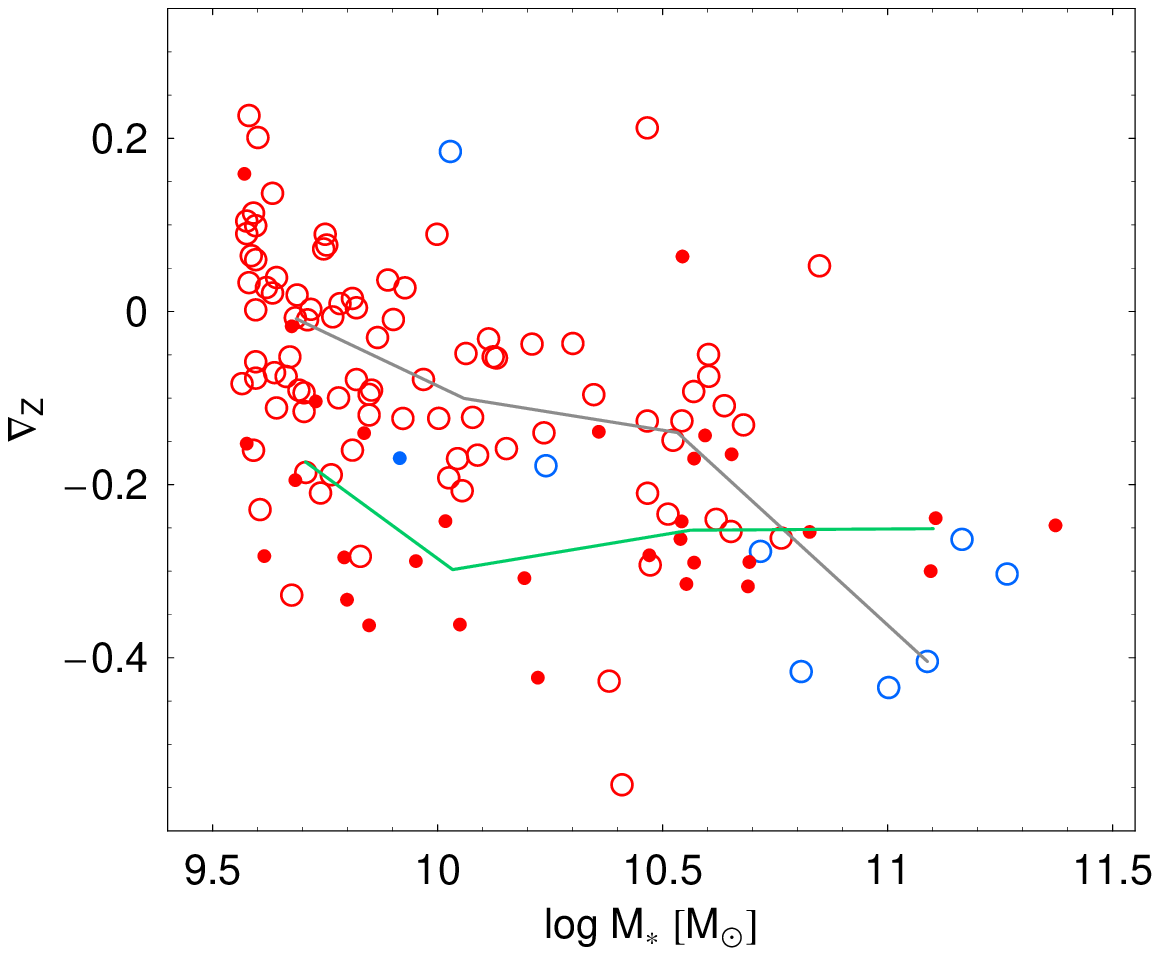, width=0.32\textwidth}\hspace{0.15cm}
\psfig{file=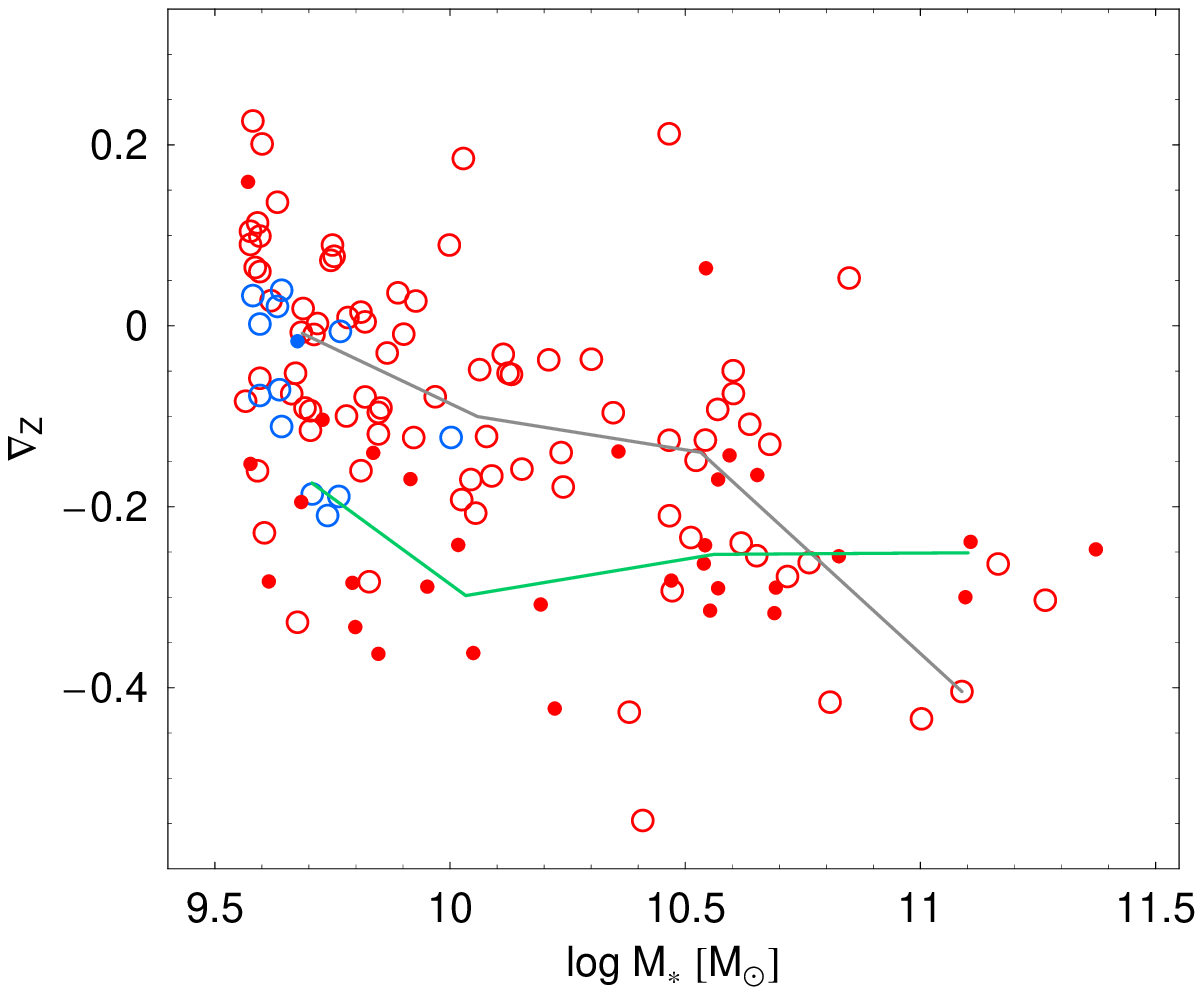, width=0.32\textwidth} \caption{Age (top
panels) and metallicity  (bottom panels) gradients as a function
of stellar mass. Filled and open symbols are for cluster and group
galaxies, respectively. Continue gray and green lines are the
medians for groups and clusters. From the left to right we divide
galaxies in bins of total colour $B-V$, specific star formation
SSFR, and velocity dispersion. {\it First panels:} Red and blue
symbols are for galaxies redder and bluer than $B-V=0.8$. {\it
Second panels:} Red and blue symbols are for galaxies with $\log
SSFR$ lower and higher than $-2$. {\it Third panels:} Red and blue
symbols are for a total velocity dispersion higher and lower than
$150 \, \rm km/s$, respectively.} \label{fig:grad_vs_mass_z0_bis}
\end{figure*}

\section{Results}\label{sec:results}

We show the age and metallicity gradients for groups and clusters
as a function of stellar mass in Fig. \ref{fig:grad_vs_mass_z0}.
As shown in the upper panels of this figure, on average, low mass
systems have null age gradients and very shallow (either negative
or positive) metallicity gradients. However, a cloud of high age
gradients in galaxies with $\log \mst \lsim 10.5 \, \Msun$ is
found in the external regions of simulated clusters and in normal
groups, while galaxies in cluster cores and fossil groups have a
quite tight distribution around $\gage \lsim 0.05$ (see bottom
panels). This is an interesting hint of the effect of the
environment on the low mass systems, acting on flattening age
gradients in higher density regions. Moreover, galaxies in the
groups have shallower metallicity gradients, when compared with
clusters. In particular, FG galaxies are distributed along a
tighter sequence in both the age and the metallicity gradients.

\begin{figure*}
\psfig{file=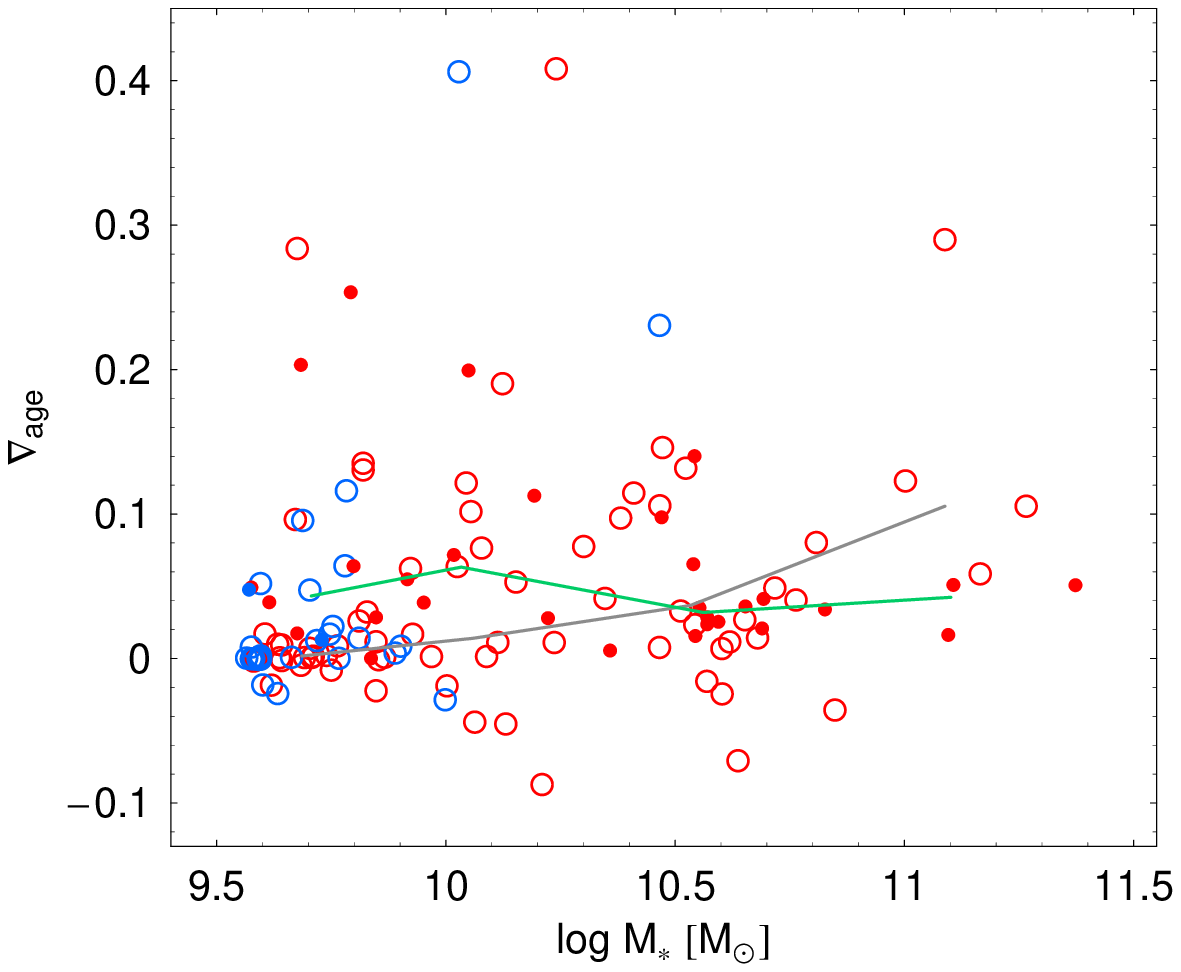, width=0.39\textwidth}\hspace{0.3cm}
\psfig{file=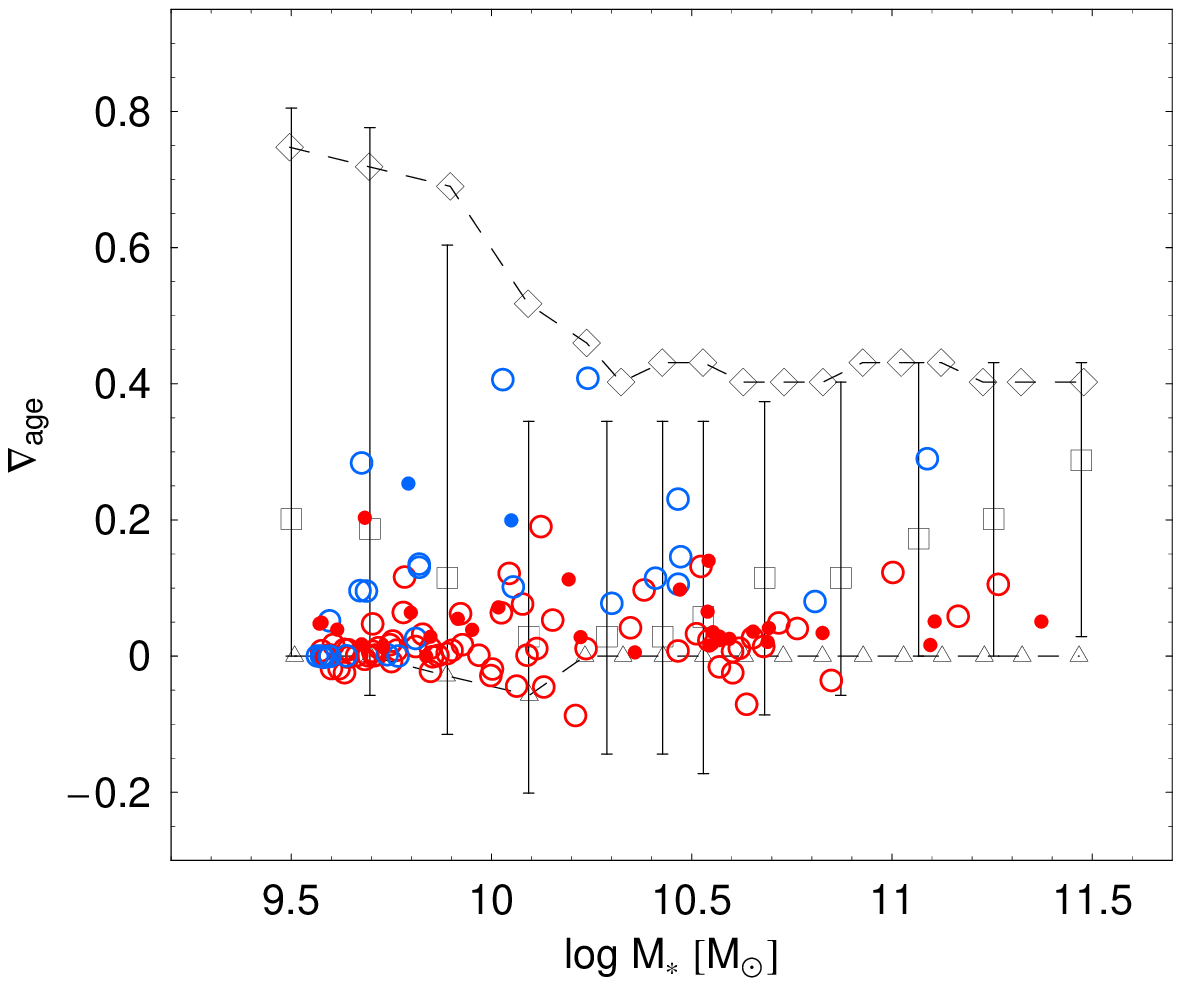, width=0.39\textwidth}\\
\psfig{file=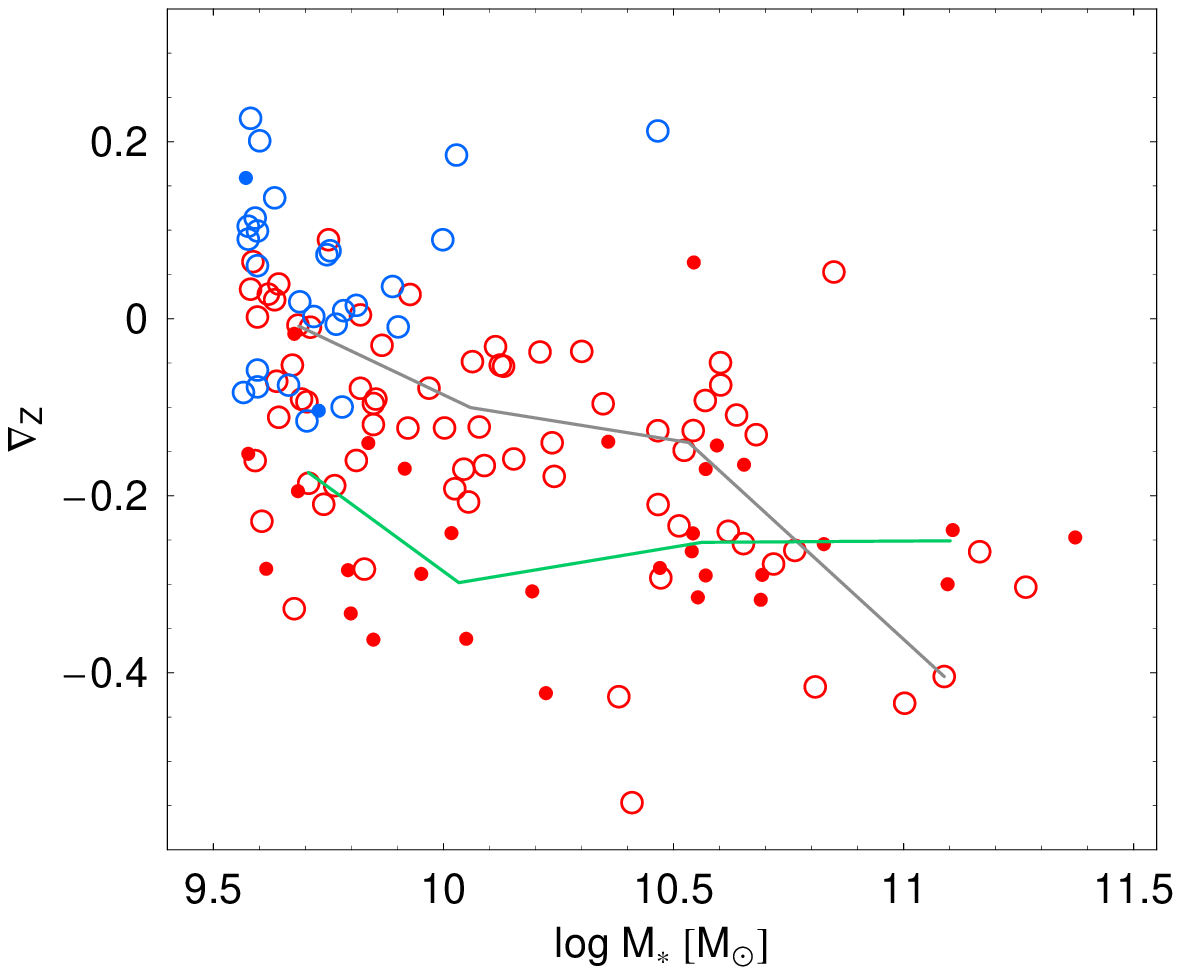, width=0.39\textwidth}\hspace{0.3cm}
\psfig{file=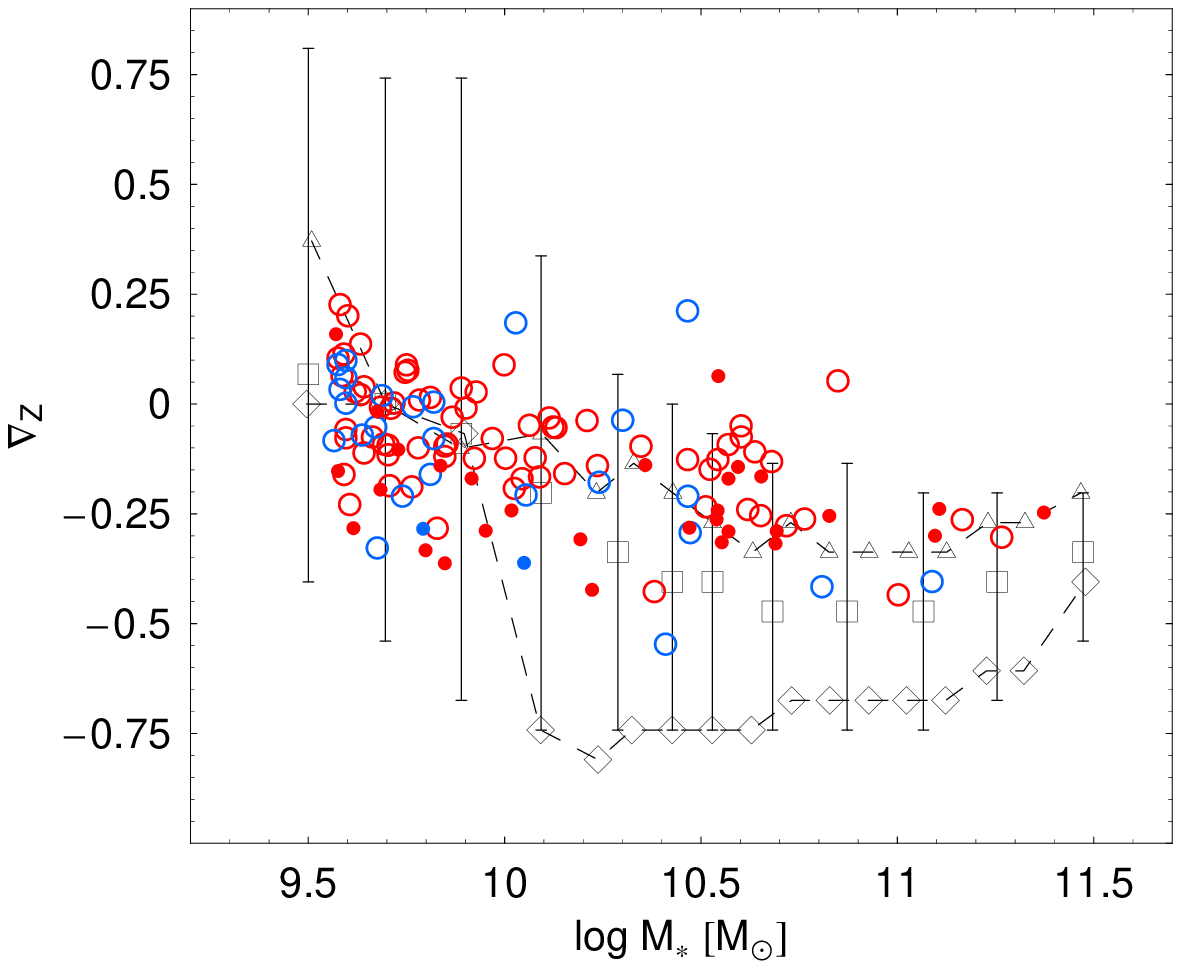, width=0.39\textwidth} \caption{Age (top
panels) and metallicity (bottom panels) gradients as a function of
stellar mass. Filled and open symbols are for cluster and group
galaxies, respectively. The continue gray and green lines are the
medians for groups and clusters. From the left to right we divide
galaxies in bins of central metallicity and age. {\it Left
panels:} Red and blue symbols are for galaxies more and less
metal-rich than $0.008$ at their centre. {\it Right panels:} Red
and blue symbols are for galaxies centrally older and younger than
$6\, \rm Gyr$. Here, we compare with results from the analysis of
the local sample of ETGs in T+10. The boxes with bars are the
results for all the sample, while the squares and triangles are
for galaxies with a central age $<6\, \rm Gyr$ and $> 6 \, \rm
Gyr$, respectively.}\label{fig:grad_vs_mass_z0_ter}
\end{figure*}

At higher masses we have found slightly positive age gradients and
negative metallicity gradients. On the side of the
most massive galaxies ($\log \mst \gsim 10.7 \Msun$), the mean age
gradients become increasingly positive with mass at the high-mass
end; on the contrary, the trend of metallicity gradients at high
masses gets flatter with mass. At these mass scales, (normal)
group galaxies show higher positive age gradients and steeper
negative metallicity gradients. In particular, while groups
present a continuous steepening of metallicity gradients with mass,
%
%and a flattening of the trend at very high masses,
%
the mass trend of cluster members turns from steep to flat at
around $\log \mst \sim 10.3-10.4$, to slightly increasing thence
on.
%
%with marginal evidences of a trend inversion at $\log \mst \gsim 10.3-10.4$.
%

Such trend inversion in the slope of
metallicity gradients is corroborated by comparing with data from
a sample of cluster and group galaxies collected in
\citealt{Spolaor09} (see also T+09) and shown in the top right
panel of Fig. \ref{fig:grad_vs_mass_z0}. Although their low mass
region is populated by cluster galaxies only, at the massive side
cluster galaxies have shallower metallicity gradients than the
ones in group galaxies, qualitatively confirming our trends. The
slope of the massive side is even steeper with mass than ours,
probably and partially due to a dearth of very massive galaxies in
our simulated sample (\citealt{Romeo+05}), apart of the BCGs.
Likewise, our cluster data follow a less steep trend
than theirs at the low mass side too.

As shown from the error bars in the upper panels of Fig.
\ref{fig:grad_vs_mass_z0}, the estimated gradients in low mass
systems are intrinsically noisier than the ones for massive
galaxies. The error bars give an estimate of the uncertainty in
the gradients and contribute to the scatter in the derived trends
with stellar mass. See \S\ref{app:app1b} for further details about
the procedure performed to obtain such results.

In the same Fig. \ref{fig:grad_vs_mass_z0} we add a comparison
with other models as well, mostly from high-resolution
hydrodynamical simulations of disk mergers (\citealt{BS99} and
\citealt{Hopkins+09a}). In particular, the {\it Z} gradients of
massive systems fall in the range covered by merger remnants in
\cite{Hopkins+09a}. Our trend for metallicity gradients is also
qualitatively consistent with the prediction of the
chemo-dynamical model by \cite{KG03}. Instead, our results are
quite inconsistent with the almost flat gradients derived from
merging model in \cite{BS99}. Finally, when comparing with
simulations in \cite{Kawata01}, we note that our points at
intermediate mass are placed along their curve derived from models
with strong stellar feedback, at least till $10^{11}M\odot$. To
this regard, in the discussion below we will come back upon the
role of SN winds in shaping the trend of $Z$ gradients with mass.

To check the dependence on other parameters than stellar mass, in
Fig. \ref{fig:grad_vs_mass_z0_bis} we discuss the trends in Fig.
\ref{fig:grad_vs_mass_z0} by splitting the sample in two classes
of galaxy colours, specific SFR over the last Gyr ($\rm SSFR =
SFR/\mst$) and velocity dispersion. At $\log \mst \lsim 10.5$ the
bluest galaxies have the highest (positive) metallicity gradients,
while this result is inverted for massive galaxies. The age
gradients do not show particular trends. In terms of SSFR, only
few galaxies are star forming at $z$=0 (see R+08), and mostly the
more massive ones in groups; these have positive age gradients and
the steepest negative metallicity gradients. Note that few of the
most massive galaxies (with $\log \mst \gsim 10.7$) in our sample,
with a recent star formation history ($\log SSFR > -2$) and bluer
colours ($B-V < 0.8$), present steep age and metallicity gradients
and are likely candidate for late-type systems. Merging and SN
feedback alone do not allow these systems to move onto the
red-sequence, requiring the inclusion of other sources of feedback
(e.g., AGN feedback) to quench SF and flatten the gradients.
Finally, we also analyze the effect of a cut in velocity
dispersion: due to the tight correlation with stellar mass, the
systems affected by this criterion are predominantly low mass
systems in groups.

In Fig. \ref{fig:grad_vs_mass_z0_ter}, galaxies are classified by
their values of central metallicity and age. We confirm that both
metallicity and age gradients depend on central quantities,
respectively (e.g., \citealt{Hopkins+09a}, \citealt{Rawle+10},
T+10). In particular, metallicity gradients strongly depend on the
central metallicity, with metal-poor systems having higher values
(null or positive). In the last panels, we can also report a
substantial agreement between the simulations and the median
gradients recovered from local ETGs in T+10 (black symbols in the
right panels of Fig. \ref{fig:grad_vs_mass_z0_ter}). As found in
T+10, here we can see that metallicity and age gradients are
strong functions of central age: massive and older systems have
null age gradients and shallower metallicity gradients ($\sim
-0.2,-0.3$). This is quite in agreement with other observations
(e.g. \citealt{Spolaor09}; \citealt{Rawle+10}) and simulations of
massive merger remnants discussed above (\citealt{BS99};
\citealt{Ko04}; \citealt{Hopkins+09a}). On the other hand, the
agreement between observed and simulated younger galaxies is
looser, what might be tracked to the absence of those field
galaxies in the simulated sample that are expected to have steeper
gradients (\citealt{LaBarbera2005}, \citealt{Tortora+10CGbis}).

\section{Discussion}\label{sec:discussions}

The net result of the observed quantities at $z$=0 is the
reflection of all the physical processes modelled in the
simulation code (namely radiative cooling, stellar winds,
supernova feedback and galaxy merging) and represents the endpoint
of galaxy evolution along the cosmic history. The next step would
be the investigation of the evolution of the population gradients
with redshift, which will be the topic of a forthcoming paper.
Even at this stage though, we can build a clear picture of the
relative contributions of each physical process in shaping the
present day trends.

The overall trend of metallicity gradients with mass is only partially
consistent with the expectation from a monolithic collapse, where
the accretion of SSPs produces a higher central metallicity and
negative metallicity gradients, that are steeper at larger masses
where the potential well is deeper.

At high masses, the null age gradients are the consequence of
galaxy merging that produces a mixing of the SSPs after the event.
The null gradients are in fact more common in cluster environment and
contrast with the presence of positive gradients in (normal) group
environment (see e.g. Fig. \ref{fig:grad_vs_mass_z0}), where there
are still a few massive star forming massive systems (Fig.
\ref{fig:grad_vs_mass_z0_bis}).
%
%following the typical monolithic-like pattern.
%
The ``merging remnant'' systems are also the ones having the
shallower (negative) metallicity gradients with respect to the
steeper ones of the star-forming systems. This shows that at the
massive side merging is still a major player of the galaxy
evolution -what makes our results consistent with former
simulations of galaxy mergers (see Fig. 1 again).

In fact, typically major dry mergers are known to level out
pre-existing metallicity and colour gradients (see
\citealt{Pipino+10}, \citealt{DiMatteo09}), hence this feature
imposes a constraint upon the relevance of merging over passive
evolution at the high mass end. In particular, at the highest
masses the occurrence of strong mass accretion due to minor and,
mainly, major mergings prevents the gradients to further steepen.

At these mass scales, the amount of energy ejected by SNe
contributes to yield steep {\it Z} gradients in massive
star-forming galaxies, which still have to experience major
merging events. On the other mass side (i.e. $\log \mst<10 \Msun$)
instead, SNe produce, on average, null gradients due to the large
meshing action required by winds easily propagating within their
smaller volume and less deep potential wells. In this case we see
that star formation has been shut off in all classes of galaxies
(Fig. \ref{fig:grad_vs_mass_z0_bis}), and the age gradients are
similarly close to zero. Environment acts as to make galaxies in
clusters have negative gradients with respect to the groups. This
suggests a different response of the enriched medium against SNe
explosions. In lower density environments, the metals are kept in
the small systems and recycled until the overall metallicity
gradients are swept out, while in the cluster core environment the
higher density and the more numerous encounters imply that part of
the metals are lost in the inter-galactic medium and the net
metallicity gradients stay negative as for massive systems.

However, there are a number of galaxies (both in cluster outskirts
and groups) which seem to diverge from the overall age trend at
the low mass end, since they show positive age gradients but
metallicity gradients consistent with the averages of the sample.
These systems do not correlate with other galaxy properties, like
the star formation, color or velocity dispersion (see e.g. Fig.
\ref{fig:grad_vs_mass_z0_bis}). Instead they are the systems with
lower ages (see Fig. \ref{fig:grad_vs_mass_z0_ter}), well matching
the suggestion that age is responsible of the scatter of the
gradients (especially the age ones) in the observed systems (see
\citealt{Tortora+10CG} and Fig. \ref{fig:grad_vs_mass_z0_ter}).
These are systems that have born in the late epochs and that have
evolved quite passively so far without experiencing any other
stellar processes but stellar ageing. We have checked that these
galaxies
%
%have been characterized by shallow metallicity gradients
%even in the past and
%
are gently migrating toward the zero gradients which they will
reach in the near future. Most of these dwarf-like systems
resulted to be star forming at $z > 0$ (R+08) and are now dead
mainly as a consequence of their interaction with close companions
(as demonstrated by the fact that the few star forming galaxies
are in normal groups and cluster outskirts), while SNe seem not to
play a major role.  At the same time these systems could not be
residing in FG and inner cluster regions because they had merged
at earlier epochs onto the brightest central galaxies.

%Another interpretation could be that these systems are not present
%in FG and inner region of cluster, since they have merged into the
%most massive galaxies, while is not the same in lower density
%environment.

\section{Conclusions}\label{sec:conclusions}

In this paper, we have used the N-body+hydrodynamical simulations
described in R+08 to study the age and metallicity gradients in
galaxies as functions of mass in the range $\log \mst \in
(9.5-11.5)$ and environment (from groups to cluster centers). Our
results show different trends of age and metallicity gradients
with mass and galactic environment at $z$=0. With respect to the
mass: 1) dwarf galaxies have generally null age gradients and
shallower or null metallicity gradients (with few cases of
positive age gradients); 2) massive early-types ($\log \mst \sim
10.5$\Msun) have moderate age gradients that slightly increase
with mass, and negative metallicity gradients ($\sim -0.2,-0.3$,
see e.g. \citealt{Rawle+10}, T+10), which reach the steepest
values $\sim -0.4$ in few of the most massive galaxies ($\log \mst
\gsim 10.5$). At fixed stellar mass there is a clear dependence on
the central galaxy age, since younger galaxies have positive age
gradients and slightly steeper $Z$ gradients, and central
metallicity. Similarly, some marginal trends are found as a
function of SSFR and colour: in particular the very massive
systems with bluer colours and active SF are found to have steeper
gradients. This mass separation in the profile slope is in
agreement with other models (e.g. \citealt{Kawata01},
\citealt{KG03}, \citealt{Hopkins+09a}), and observations
(\citealt{Rawle+10}, \citealt{Spolaor09}, \citealt{Tortora+10CG}),
even though the statistic for our very massive side is quite poor.

Such behaviour can be explained in terms of the different role
played by merging. At $\log \mst \lsim 10.5$ mergers are rare and
thus less effective in mixing stellar population (e.g.
\citealt{deLucia06}); then the only physical mechanisms in action
are the SN feedback and the passive ageing (similarly to a
monolithic collapse scenario), that result into the decreasing
trend of $Z$ gradients (see also \citealt{KG03},
\citealt{Pipino+10}).

On the other side, the flattening of metallicity gradients trend
at larger masses can be due to the higher frequency of major dry
mergers at low redshift, which have an increased efficiency in
producing flatter metallicity profiles. \cite{Pipino+10} have
demonstrated that equal mass dry mergers between ellipticals
systematically halve the slope of any pre-existing metallicity
gradient.

With respect to the environment, systems with positive age
gradients tend to be found in the external parts of clusters and
in groups, \emph{independently of galaxy's mass} and with a quite
large spread; while galaxies in the cluster cores and fossil
groups have, on average, quasi-null age gradients, again at all
masses. However, when analyzing the average values, cluster dwarf
galaxies have, steeper metallicity gradients with respect to the
ones in groups, which also present shallower values more peaked
around zero (with a larger fraction of positive gradients); slight
steeper age gradients are observed in cluster galaxies too. At
very high mass, albeit the lower statistics, this trend gets
inverted and galaxies in groups have steeper (negative)
metallicity gradients (\citealt{LaBarbera2005}). As to the scatter
of the relations, cluster cores and FGs present the tightest
distribution of age gradients, while the metallicity gradients
show no strong differences in the scatter among the different
environments.

Therefore, besides the main role of SN feedback, we have found
that environment shapes differently the gradients at low and high
masses. At low masses, tidal interactions are acting to give
steeper metallicity gradients in cluster galaxies, while merging
produces shallower gradients in massive systems in the same
environment.

In forthcoming works, we plan to extend this analysis to higher
$z$ and as well to the BCGs, to put on a firmer ground our
physical interpretations. In particular, we aim at further
investigating the physical processes that rule the very massive
galaxies such as galaxy major merging and AGN feedback
(\citealt{AS08}, \citealt{Tortora2009AGN}), which would be
important to shape both the global properties and the gradients in
stellar populations.

\section*{Acknowledgments}

We thank the anonymous referee for his suggestions which helped to
improve the paper. CT was funded by the Swiss National Science
Foundation. ADR acknowledges support from ALMA-CONICYT FUND
through grant 31070023, from FONDECYT - Proyecto de Iniciaci\'on a
la Investigaci\'on No. 11090389 and from UNAB - Proyecto Regular
No. DI-35-09/R. AM acknowledges partial support from Proyecto
Regular de Investigacion UNAB DI-41-10/R.

\appendix

\section{Systematics}\label{app:syst}

In this section we will analyze various systematics induced by
selection criteria and the fit procedure adopted to derive the age
and metallicity gradients.

\subsection{Fitting procedure}\label{app:app1a}

\begin{figure*}
\psfig{file=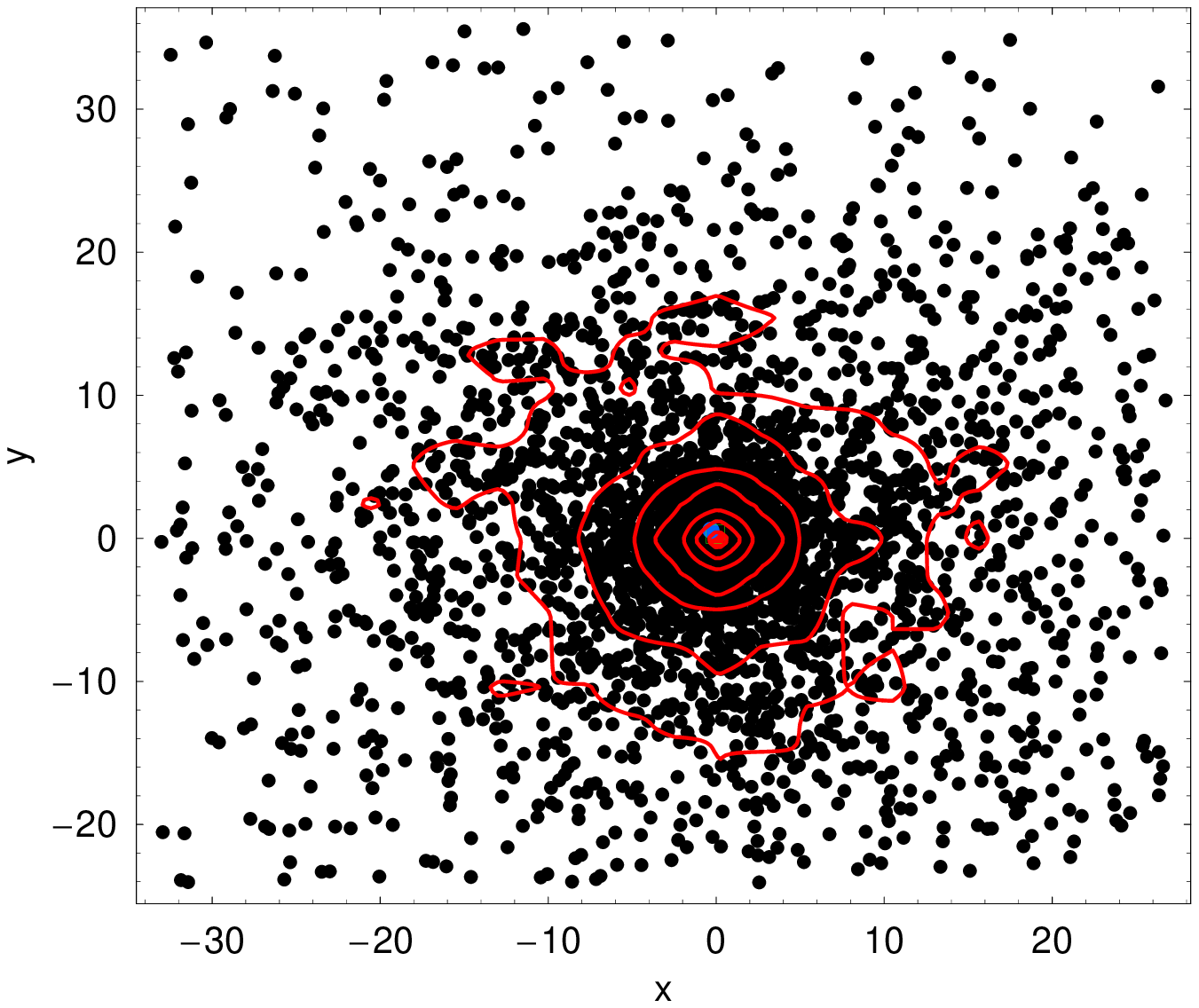, width=0.33\textwidth}
\psfig{file=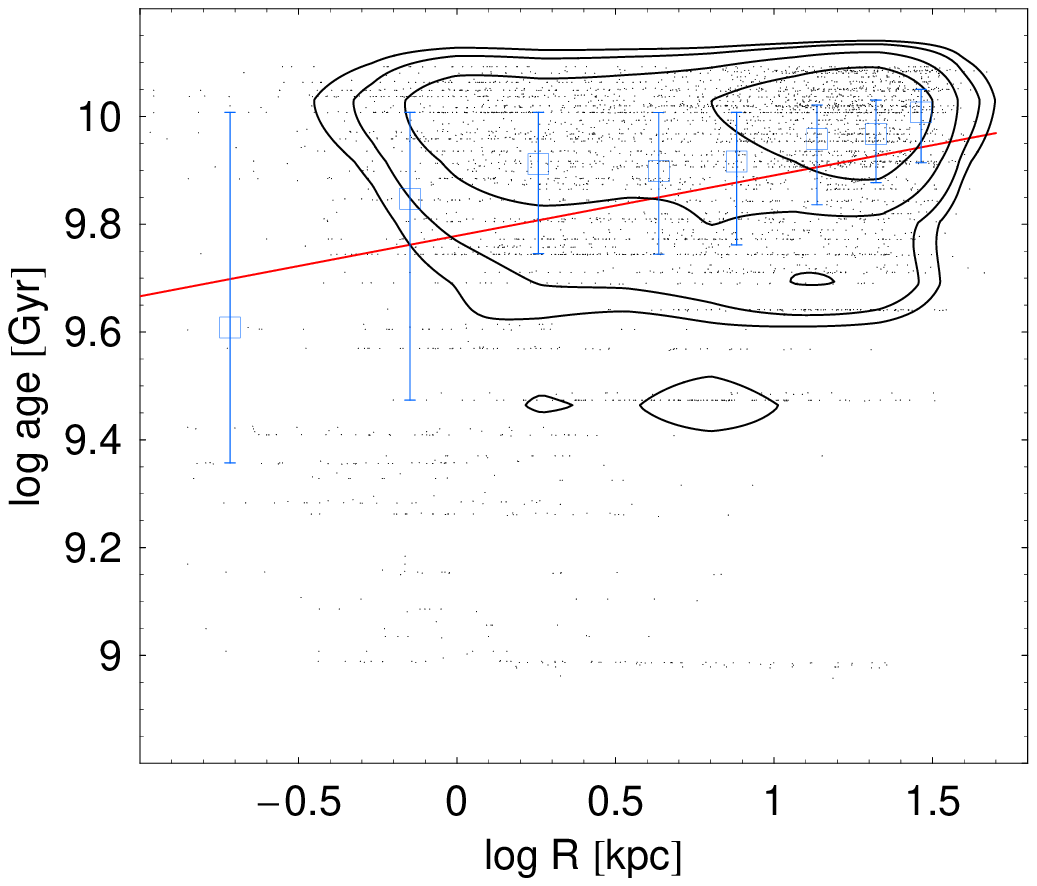, width=0.33\textwidth}
\psfig{file=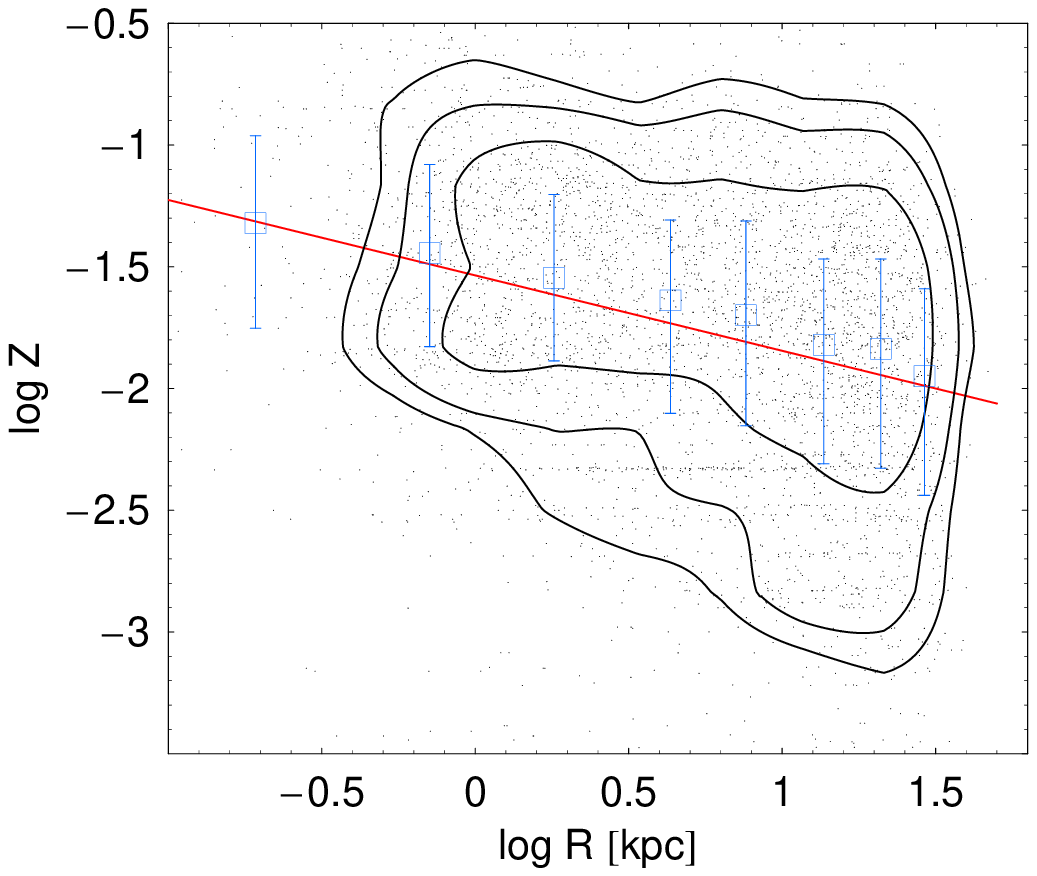, width=0.33\textwidth}\\
\psfig{file=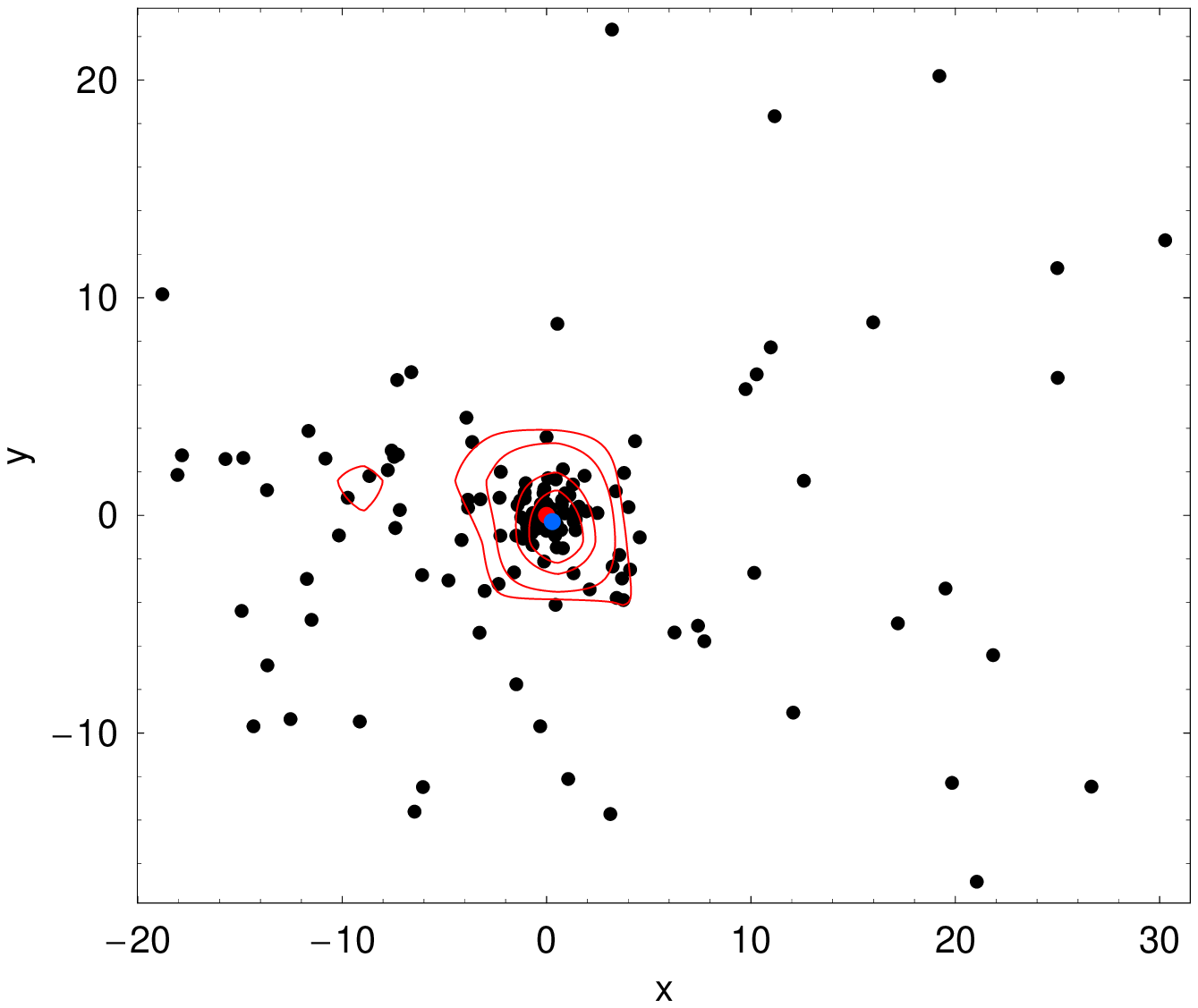, width=0.33\textwidth}
\psfig{file=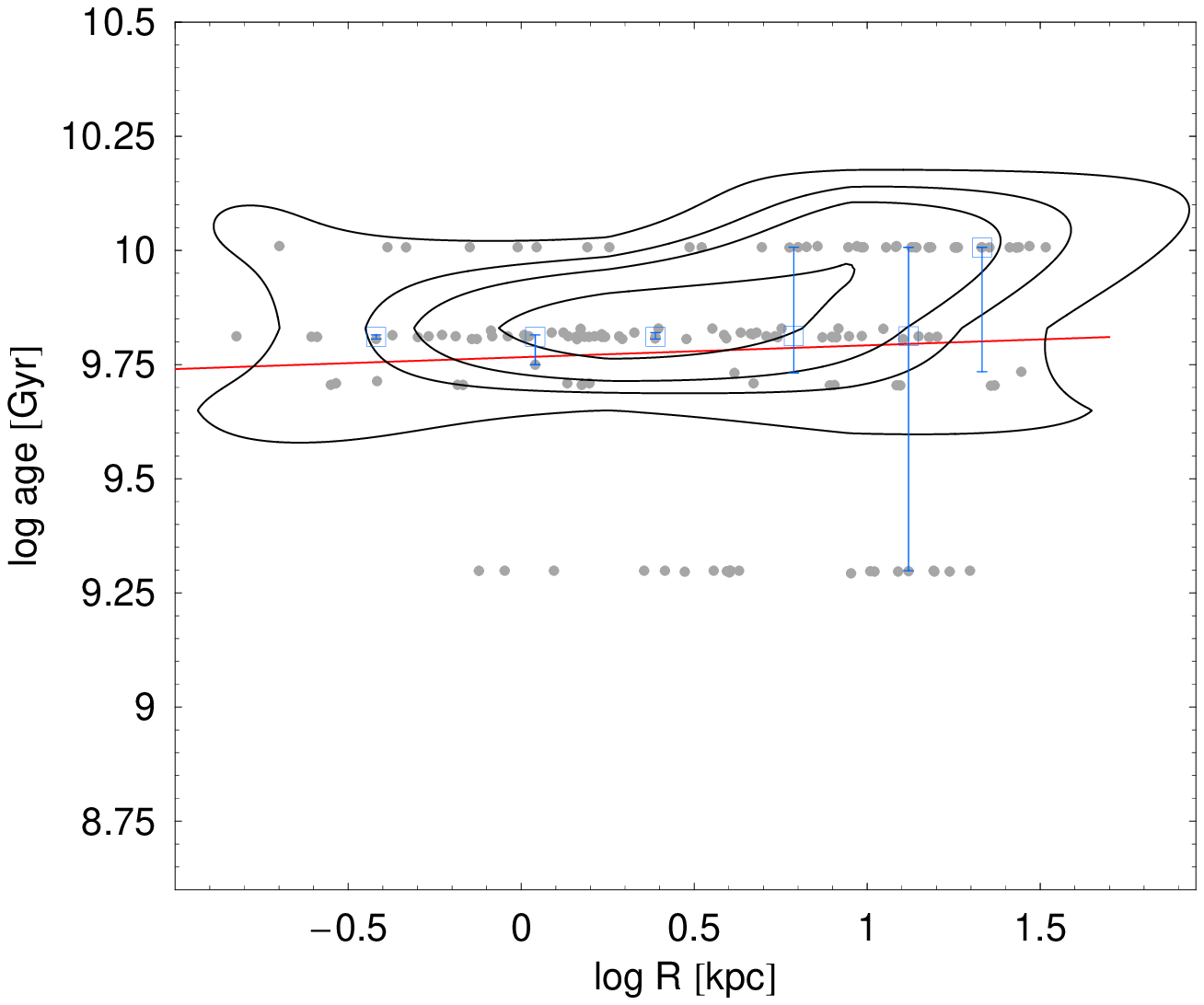, width=0.33\textwidth}
\psfig{file=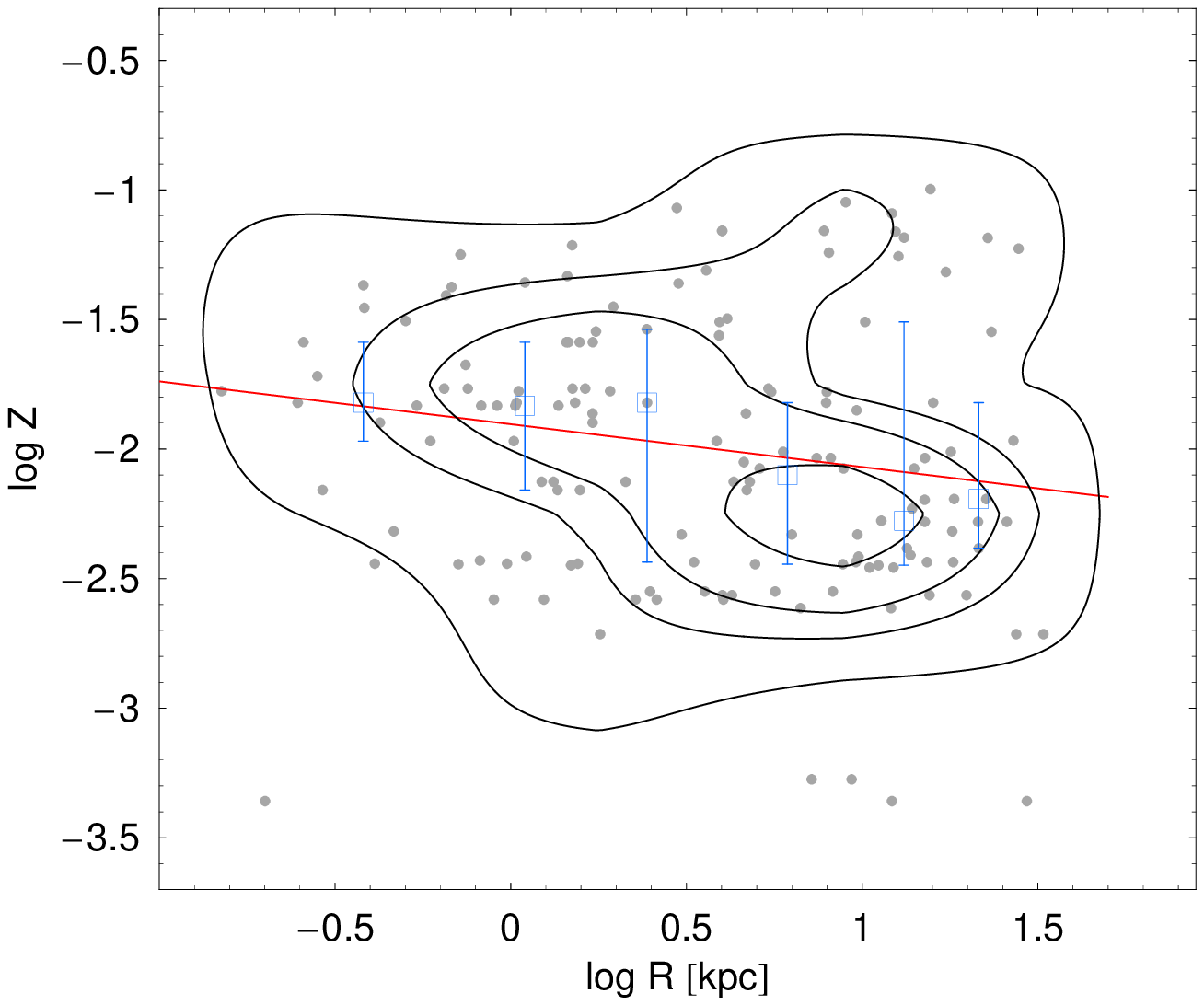, width=0.33\textwidth} \caption{Particle
distributions, age and metallicity profiles from the left to
right, for a massive  (top panels) and dwarf (bottom panels) group
galaxy. {\it Left panels.} Distribution of particles in the plane
x$-$y. The axes coordinates are re-normalized to the the median of
the particles (red point), while the blue point is the mean. The
red contours show the density of data-points. {\it Middle panels.}
Galaxy age profile shown as $\log age$ vs $\log R$, the contours
show the density of data-points, the blue bars are the medians
with the 25-75th percentiles, while the red line is the linear fit
we have performed. {\it Right panels.} The same as the middle
panels but for the metallicity profile.}\label{fig:fig_app_1}
\end{figure*}

As described in the text, we have performed a linear fit $\log X -
\log R$ in order to recover the age and metallicity profiles of
the simulated galaxies. The first basic step of such a procedure
is to derive the galaxy centre from the distribution of the star
particles. The particle distributions for two template galaxies
are shown in Fig. \ref{fig:fig_app_1}. The choice of the median as
the galaxy centre is dictated by the scarce sensitivity of such
estimator to the outliers: in many cases indeed, the particle
distribution is not symmetric (as shown in the bottom panel of
Fig. \ref{fig:fig_app_1}), therefore the median is more suitable
to determine the minimum of the potential well.

The profiles for galaxy age and metallicity are shown in Fig.
\ref{fig:fig_app_1} too. The data points are compared with our
best fitted profile, iso-density contours and medians in different
radial bins. The fitting procedure is directly performed on the
collection of points, cutting the tails of age and metallicity
distributions out of 3$\sigma$, to avoid systematics from outliers
affecting the slope's estimate.

Finally, we have also analyzed the change in the derived gradients
when the age and metallicity profiles are fitted in a limited
radial range, as made with observations. We fit the profiles using
only the particles in the range $(\Re/10, \Re)$, where $\Re$ is
the radius which encloses half of the stellar mass, and the final
results are qualitatively unchanged.

\begin{figure*}
\psfig{file=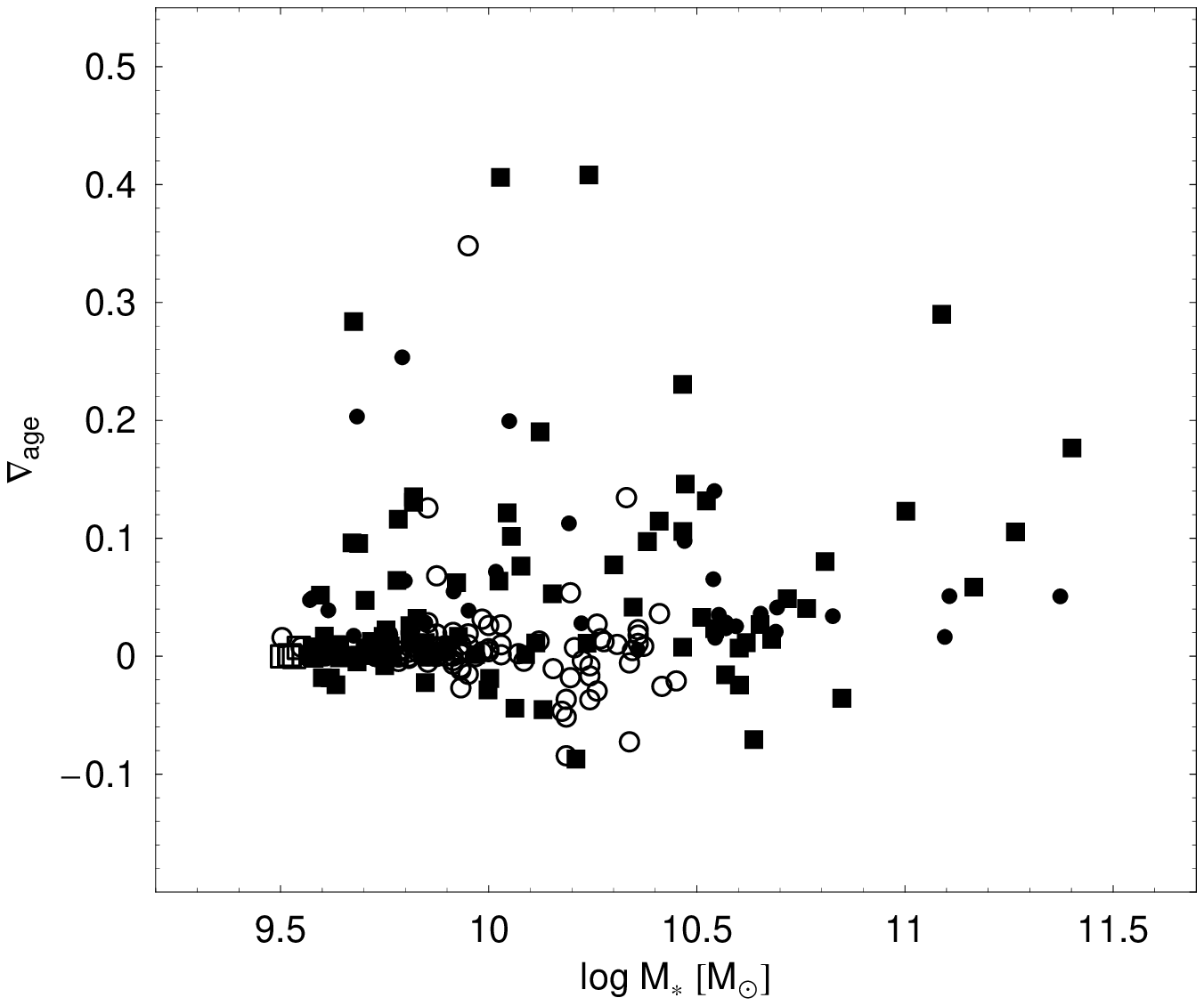, width=0.45\textwidth}
\psfig{file=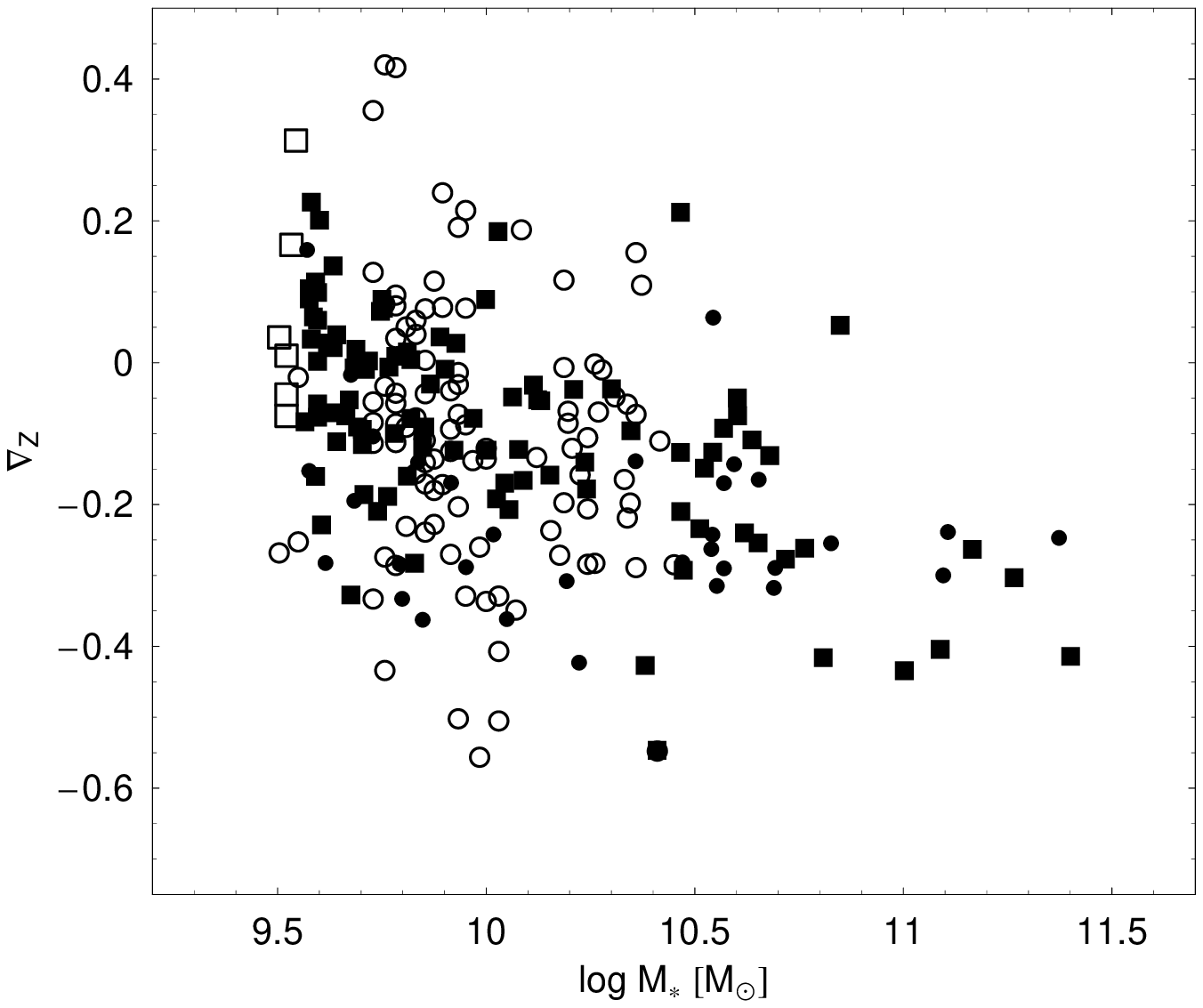, width=0.45\textwidth} \caption{Age (left
panel) and metallicity (right panel) gradients as a function of
stellar mass. Cirles and boxes are for cluster and group members,
respectively. Filled and open symbols are for galaxies with
$N_{par}>80$ and $N_{par}\leq 80$,
respectively.}\label{fig:fig_app_2}
\end{figure*}

\subsection{Systematics induced by the small number of
particles}\label{app:app1b}

We have performed a series of tests on our galaxies to understand
what is the spurious scatter introduced in fitting objects
with a very low number of particles. We have taken some galaxies
with a high number of particles (in the range $1000-5000$) and
fixed the derived slopes as with a null uncertainty. Then we
have extracted from them $1000$ synthetic galaxies having a number
of particles $N_{par} = 70, \, 225, \, 710$ and when possible
$=2250$, which correspond, for the highest resolution in our
simulations, to $\log \mst \sim 9.5, \, 10, \, 10.5$ and $11.5$,
respectively. The fitting procedure is performed on each of these
1000 synthetic systems  and for each $N_{par}$; from the
slopes hence obtained we have derived a best median value and a $1 \sigma$
uncertainty. While the best fitted slope of the original galaxies
are perfectly recovered, with an almost null scatter, the
uncertainty is not negligible and is, obviously, larger for very
low mass systems, while is very small for the most massive
galaxies with $\log \mst \sim 11$. We have presented these results
in Fig. \ref{fig:grad_vs_mass_z0}.

As already discussed, this uncertainty can produce part of the
larger scatter observed in the gradients at very low masses and
these systematics could be stronger when the number of particles
is lower than the minimum value of $80$ we have imposed. To
understand the impact on our results, in Fig. \ref{fig:fig_app_2}
we show the age and metallicity gradients for both group and
cluster members as a function of mass, classifying them on the
basis of the number of particles. The cluster members are the
systems strongly affected by our selection criterion on $N_{par}$,
that cuts out many galaxies with $\log \mst \lsim 10.5$. Such a
cut reduces the spread in the metallicity gradients, while for the
age gradients no relevant variations are reported.


\begin{thebibliography}{99}

\bibitem[\protect\citeauthoryear{Antonuccio-Delogu et al.}{2003}]{AD+03} Antonuccio-Delogu V., Becciani U., Ferro D., Romeo A., 2003, Mem.
Soc. Astron. Ital. Suppl., 1, 109

\bibitem[\protect\citeauthoryear{Antonuccio-Delogu \& Silk}{2008}]{AS08} Antonuccio-Delogu V.
\& Silk J. 2008, MNRAS, 389, 1750

\bibitem[\protect\citeauthoryear{Arimoto \& Yoshii}{1987}]{AY87} Arimoto N. \& Yoshii Y. 1987, A\&A, 173, 23



\bibitem[\protect\citeauthoryear{Bekki \& Shioya}{1999}]{BS99} Bekki, K., \& Shioya, Y. 1999, ApJ, 513, 108

\bibitem[\protect\citeauthoryear{Benson et al.}{2001}]{Benson+01}  Benson A.J., Pearce F.R., Frenk C.S., Baugh C.M., Jenkins A.,
2001, MNRAS, 320, 261

\bibitem[\protect\citeauthoryear{Carlberg}{1984}]{Carlberg84} Carlberg R. G., 1984, ApJ, 286, 403

\bibitem[\protect\citeauthoryear{Carollo et al.}{1993}]{CDB93} Carollo C. M., Danziger I. J. \& Buson L., 1993,
MNRAS, 265, 553


\bibitem[\protect\citeauthoryear{Chabrier }{2001}]{Chabrier01} Chabrier, G. 2001, ApJ, 554, 1274

\bibitem[\protect\citeauthoryear{Davies et al.}{1993}]{Davies+93} Davies R. L., Sadler E. M., Peletier R. F. 1993, MNRAS, 262, 650


\bibitem[\protect\citeauthoryear{de Lucia et al. }{2006}]{deLucia06} de Lucia, G., Springel, V., White, S. D. M., Croton, D.,
Kauffmann, G. 2006, MNRAS, 366, 499D



\bibitem[\protect\citeauthoryear{Di Matteo et al.}{2009}]{DiMatteo09} di Matteo P., Pipino A., Lehnert M. D., Combes F., Semelin B.
2009, A\&A, 499, 427

\bibitem[\protect\citeauthoryear{Forbes et al.}{2005}]{Forbes+05} Forbes D.A. S$\rm\acute{a}$nchez-Bl$\rm\acute{a}$zquez P. \&
Proctor R. 2005, MNRAS, 361, 6



\bibitem[\protect\citeauthoryear{Gibson}{1997}]{Gibson97} Gibson B. K. 1997, MNRAS, 290, 471


\bibitem[\protect\citeauthoryear{Helly et al.}{2002}]{Helly+03} Helly J. C., Cole S., Frenk C. S., Baugh C. M., Benson A., Lacey
C., Pearce F. R., 2003, MNRAS, 338, 913

\bibitem[\protect\citeauthoryear{Hopkins et al.}{2009a}]{Hopkins+09a} Hopkins P. F., Cox T. J., Dutta S. N., Hernquist L., Kormendy
J., \& Lauer T. R. 2009a, ApJS, 181, 135


\bibitem[\protect\citeauthoryear{Kawata}{2001}]{Kawata01} Kawata D. 2001, ApJ, 558, 598


\bibitem[\protect\citeauthoryear{Kawata \& Gibson}{2003}]{KG03} Kawata D. \& Gibson B. K. 2003, MNRAS, 340, 908


\bibitem[\protect\citeauthoryear{Kobayashi \& Arimoto} {1999}]{KoAr99} Kobayashi C. \& Arimoto N. 1999, ApJ, 527, 573

\bibitem[\protect\citeauthoryear{Kobayashi} {2004}]{Ko04} Kobayashi C. 2004, MNRAS, 347, 740


\bibitem[\protect\citeauthoryear{La Barbera et al.} {2005}]{LaBarbera2005}  La Barbera F. et al. 2005, ApJ, 626, 19


\bibitem[\protect\citeauthoryear{Larson}{1974}]{Larson74} Larson R. B., 1974, MNRAS, 166, 585


\bibitem[\protect\citeauthoryear{Larson}{1975}]{Larson75} Larson R. B., 1975, MNRAS, 173, 671


\bibitem[\protect\citeauthoryear{Mihos \& Hernquist}{1994}]{MH94} Mihos J. C. \& Hernquist L., 1994, ApJ, 437, L47


\bibitem[\protect\citeauthoryear{Peletier et~al.}{1990a}]{Peletier+90a} Peletier R.F., Davies R.L., Illingworth G.D., Davis L.E., \&
Cawson, M. 1990, AJ, 100, 1091


\bibitem[\protect\citeauthoryear{Peletier et~al.}{1990b}]{Peletier+90b} Peletier R.F., Valentijn E.A., \& Jameson, R.F. 1990, A\&A, 233,
62


\bibitem[\protect\citeauthoryear{Peletier et al.}{2007}]{Peletier+07} Peletier R.F. et al. 2007, MNRAS, 379, 445


\bibitem[\protect\citeauthoryear{Pipino et al.}{2008}]{Pipino08} Pipino A., D'Ercole A. \& Matteucci F. 2008, A\&A, 484, 679

\bibitem[\protect\citeauthoryear{Pipino et al.}{2010}]{Pipino+10} Pipino A., D'Ercole A., Chiappini C. \& Matteucci F. 2010, arXiv:1005.2154

\bibitem[\protect\citeauthoryear{Rawle et al.}{2010}]{Rawle+10} Rawle T. D., Smith R. J. \& Lucey J. R. 2010, MNRAS, 401, 852

\bibitem[\protect\citeauthoryear{Romeo et al.}{2005}]{Romeo+05} Romeo A.D., Portinari L., Sommer-Larsen J., 2005,
MNRAS, 361, 983

\bibitem[\protect\citeauthoryear{Romeo et al.}{2006}]{Romeo+06} Romeo A.D., Sommer-Larsen J., Portinari L., Antonuccio-Delogu V., 2006, MNRAS, 371, 548



\bibitem[\protect\citeauthoryear{Romeo et al.}{2008}]{Romeo+08} Romeo A.~D., Napolitano
N.~R., Covone G., Sommer-Larsen J., Antonuccio-Delogu V., \&
Capaccioli M. 2008, MNRAS, 389, 13 (R+08)

\bibitem[\protect\citeauthoryear{Ruszkowski \& Springel}{2009}]{Ruszkowski+09} Ruszkowski M. \& Springel V. 2009, ApJ, 696, 1094


\bibitem[\protect\citeauthoryear{Sijacki et al.}{2007}]{Sijacki+07} Sijacki D. et al. 2007, MNRAS, 380, 877

\bibitem[\protect\citeauthoryear{Spolaor et al.}{2009}]{Spolaor09} Spolaor, M., Proctor, R. N., Forbes, D. A., Couch,
W. J. 2009, ApJ, 691, 138

\bibitem[\protect\citeauthoryear{Springel et al.}{2005}]{Springel05} Springel V., Di Matteo T. \& Hernquist L. 2005, ApJ, 620, 79


\bibitem[\protect\citeauthoryear{Tamura et al.}{2000}]{T+00} Tamura, N., et al. 2000, AJ, 119, 2134

\bibitem[\protect\citeauthoryear{Tamura \& Ohta}{2000}]{TO2000} Tamura, N., \& Ohta, K. 2000, AJ, 120, 533


\bibitem[\protect\citeauthoryear{Tamura \& Ohta}{2003}]{TO2003} Tamura, N., \& Ohta, K. 2003, AJ, 126, 596


\bibitem[\protect\citeauthoryear{Tortora et al.}{2009}]{Tortora2009AGN} Tortora C. et al. 2009, MNRAS, 396, 61



\bibitem[\protect\citeauthoryear{Tortora et al.}{2010}]{Tortora+10CG} Tortora C. et al. 2010, accepted on MNRAS, arXiv:1004.4896 (T+10)

\bibitem[\protect\citeauthoryear{Tortora et al.}{2010b}]{Tortora+10CGbis} Tortora C. et al. 2010b in preparation


\end{thebibliography}
\end{document}